\documentstyle[preprint,aps]{revtex}
\newcommand{\f}{\begin{equation}}
\newcommand{\e}{\end{equation}}
\newcommand{\n}{\eqnum}
\begin{document}
\title{Spinors in Higher Dimensional \\ and Locally Anisotropic Spaces}
\draft
\author{Sergiu I. Vacaru}
\address{ Institute of Applied Physics, Academy of Sciences, \\
5 Academy str., Chi\c sin\v au 2028,
 Republic of Moldova. \\ E-mail:  lises@cc.acad.md}
\maketitle

\begin{abstract}
The theory of spinors is developed for locally anisotropic (la) spaces,
in brief la-spaces, which in general are modeled as vector bundles provided
with nonlinear and distinguished connections and metric structures (such
la-spaces contain as particular cases the Lagrange, Finsler and,
for trivial nonlinear connections, Kaluza-Klein spaces). The
la-spinor differential geometry is constructed. The distinguished spinor
connections are studied and compared with similar ones on la-spaces. We
derive the la-spinor expressions of curvatures and torsions and analyze the
conditions when the distinguished torsion and nonmetricity tensors can be
generated from distinguished spinor connections. The dynamical equations for
gravitational and matter field la-interactions are formulated.

{\bf \copyright S.I. Vacaru}
\end{abstract}

\pacs{PACS numbers: 04.50.+h, 04.20.Cv, 02.90.+p, 11.90.+t, 04.90.+e}
\vskip40pt
{\large \bf 1.\ Introduction}
\vskip20pt
Some of approaches to fundamental problems in physics advocate the extension
to locally anisotropic backgrounds of physical theories [1-5]. In order to
construct physical models on la-spaces it is necessary a corresponding
generalization of the spinor theory. Spinor variables and interactions of
spinor fields on Finsler spaces were used in a heuristic manner, for
instance, in works [5,6], where the problem of a rigorous definition of
la-spinors for la-spaces was not considered. Here we note that, in general,
the nontrivial nonlinear connection and torsion structures and possible
incompatibility of metric and connections makes the solution of the
mentioned problem very sophisticate. The geometric definition of la-spinors
and a detailed study of the relationship between Clifford, spinor and
nonlinear and distinguished connections structures in vector bundles,
generalized Lagrange and Finsler spaces are presented in [7,8].

The purpose of the paper is to continue our investigations [7-10] on
formulation of the theory of classical and quantum field interactions on
la-spaces. We receive primary attention to the development of the necessary
geometric framework: to propose an abstract spinor formalism and formulate
the differential geometry of la-spaces (the second step after the definition
of la-spinors in [7,8]). The next step is the investigation of locally
anisotropic interactions of fundamental fields on generic la-spaces.

For our considerations on the la-spinor theory it will be convenient to
generalize the Penrose and Rindler abstract index formalism [11-13] (see
also the Luehr and Rosenbaum index free methods [14]) proposed for
spinors on locally isotropic spaces. We note that in order to formulate the
locally anisotropic physics usually we have  dimensions $d>4$ for
the fundamental la-space-time. In this case the 2-spinor calculus does not
play a preferential role.

The presentation is divided into six sections. It begins with a necessary
reformulation of the Miron and Ansatesiei [2] geometry of la-spaces in
section 2. The spinor technique for distinguished vector spaces is
developed in section 3. The distinguished spinor connections, torsions and
curvatures and theirs relations with similar geometric objects introduced in
the framework of d-tensor formalism are defined and studied in details in
section 4. In the next section 5 we consider the d-spinor and d-tensor
formulations of the theory of classical field interactions on la-spaces.
 Outlook and discussion are contained in section 6.

\vskip60pt
{ \large \bf 2. \ Connection and Metric Structures  }

{\large \bf \qquad in Locally Anisotropic Spaces}
\vskip20pt
As a preliminary to a discussion of la-spinor formalism we summarize some
 results
and methods of the differential geometry of vector bundles provided with
nonlinear and distinguished connections and metric structures [2].\ This
section serves the twofold purpose of establishing of abstract index
denotations and starting the geometric backgrounds which are used in the next
sections of the paper.
\vskip50pt
{\large \em 2.1. \ Nonlinear connection and distinguishing }

{\large \em \qquad  of geometric objects in fibre bundles}
\vskip25pt
Let ${\cal E=}$ $\left( E,p,M,Gr,F\right) $ be a locally trivial vector
bundle, v-bundle, where $F={\cal R }^m$ (a real vector space of dimension $%
m,\dim F=m,$ ${\cal R \ }$ denotes the real number field) is the typical
fibre, the structural group is chosen to be the group of automorphisms of $%
{\cal R }^m$ , i.e. $Gr=GL\left( m,{\cal R }\right) ,\,$ and $p:E\rightarrow M$
is a differentiable surjection of a differentiable manifold $E$ (total
space, $\dim E=n+m)$ to a differentiable manifold $M$ $\left( \mbox{base
space, }\dim M=n\right) .$ Local coordinates on ${\cal E}$ are denoted as $%
u^{{\bf \alpha }}=\left( x^{{\bf i}},y^{{\bf a\ }}\right) ,$ or in brief $%
{\bf u=\left( x,y\right) ,}$ where boldfaced indices will be considered as
coordinate ones for which the Einstein summation rule holds (Latin indices $%
{\bf i,j,k,...}=1,2,...,n$ will parametrize coordinates of geometrical
objects with respect to a base space $M,$ Latin indices ${\bf a,b,c,...=}%
1,2,...,m$ will parametrize fibre coordinates of geometrical objects and
Greek indices ${\bf \alpha ,\beta ,\gamma ,...}$ are considered as
cumulative ones for coordinates of objects defined on the total space of a
v-bundle). We shall correspondingly use abstract indices $\alpha =(i,a),$ $%
\beta =(j,b),\gamma =(k,c),...$ in the Penrose manner [11-13] in order to
mark geometical objects and theirs (base, fibre)-components or, if it will
be convenient, we shall consider boldfaced letters (in the main for pointing
to the operator character of tensors and spinors into consideration) of type
${\bf A\equiv }A=\left( A^{(h)},A^{(v)}\right) {\bf ,b=}\left(
b^{(h)},b^{(v)}\right) {\bf ,...,R,\omega ,\Gamma ,...}$ for geometrical
objects on ${\cal E}$ and theirs splitting into horizontal (h), or base, and
vertical (v), or fibre, components. For simplicity, we shall prefer writing
out of abstract indices instead of boldface ones if this will not
give rise to ambiguities.

Coordinate transforms $u^{{\bf \alpha ^{\prime }\ }}=u^{{\bf \alpha ^{\prime
}\ }}\left( u^{{\bf \alpha }}\right) $ on ${\cal E}$ are written as $\left(
x^{{\bf i}},y^{{\bf a}}\right) \rightarrow \left( x^{{\bf i^{\prime }\ }},y^{%
{\bf a^{\prime }}}\right) ,$ where $x^{{\bf i^{\prime }\ }}=$ $x^{{\bf %
i^{\prime }\ }}(x^{{\bf i}}),\quad y^{{\bf a^{\prime }\ }}=M_{{\bf a\ }}^{%
{\bf a^{\prime }}}(x^{{\bf i\ }})y^{{\bf a}}$ and matrix $M_{{\bf a\ }}^{%
{\bf a^{\prime }}}(x^{{\bf i\ }})\in GL\left( m,{\cal R}\right) $ are
functions of necessary smoothness class.

A local coordinate parametrization of ${\cal E}$ naturally defines a
coordinate basis%
\f
\frac \partial {\partial u^\alpha }=\left( \frac \partial {\partial
x^i},\frac \partial {\partial y^a}\right) , \n{2.1}
\e
in brief we shall write $\partial _\alpha =(\partial _i,\partial _a),$ and
the reciprocal to (2.1) coordinate basis
\f
du^\alpha =(dx^i,dy^a), \n{2.2}
\e
or, in brief, $d^\alpha =(d^i,d^a),$ which is uniquely defined from the
equations%
$$
d^\alpha \circ \partial _\beta =\delta _\beta ^\alpha ,
$$
where $\delta _\beta ^\alpha $ is the Kronecher symbol and by ''$\circ $$"$
we denote the inner (scalar) product in the tangent bundle ${\cal TE.}$

The concept of {\bf nonlinear connection,} in brief, N-connection, is
fundamental in the geometry of la-spaces (see a detailed study and basic
references in [2]). In a v-bundle ${\cal E}$ it is defined as a distribution
$\{N:E_u\rightarrow H_uE,T_uE=H_uE\oplus V_uE\}$ on $E$ being a global
decomposition, as a Whitney sum, into horizontal,${\cal HE,\ }$ and
vertical, ${\cal VE,}$ subbundles of the tangent bundle ${\cal TE:}$

\f
{\cal TE=HE\oplus VE.\ } \n{2.3}
\e
Locally a N-connection in ${\cal E}$ is given by its components $N_{{\bf i}%
}^{{\bf a}}({\bf u})=N_{{\bf i}}^{{\bf a}}({\bf x,y})$ (in brief we shall
write $N_i^a(u)=N_i^a(x,y)$ ) with respect to bases (2.1) and (2.2)):

$$
{\bf N}=N_i^a(u)d^i\otimes \partial _a.
$$

We note that a linear connection in a v-bundle ${\cal E}$ can be considered
as a particular case of a N-connection when $N_i^a(x,y)=K_{bi}^a\left(
x\right) y^b,$ where functions $K_{ai}^b\left( x\right) $ on the base $M$
are called the Christoffel coefficients.

To coordinate locally geometric constructions with the global splitting of $%
{\cal E}$ defined by a N-connection structure, we have to introduce a
locally adapted basis (la-basis, la-frame):

\f
\frac \delta {\delta u^\alpha }=\left( \frac \delta {\delta x^i}=\partial
_i-N_i^a\left( u\right) \partial _a,\frac \partial {\partial y^a}\right) ,
\n{2.4} \e
or, in brief,$\delta _\alpha =\left( \delta _i,\partial _a\right) ,$ and its
dual la-basis%
\f
\delta u^\alpha =\left( \delta x^i=dx^i,\delta y^a+N_i^a\left( u\right)
dx^i\right) , \n{2.5}
\e
or, in brief,{\bf \ }$\ \delta \ ^\alpha =$ $\left( d^i,\delta ^a\right) .$

The{\bf \ nonholonomic coefficients }${\bf w}=\{w_{\beta \gamma }^\alpha
\left( u\right) \}$ of la-frames are defined as%
\f
\left[ \delta _\alpha ,\delta _\beta \right] =\delta _\alpha \delta _\beta
-\delta _\beta \delta _\alpha =w_{\beta \gamma }^\alpha \left( u\right)
\delta _\alpha .\n{2.6}
\e

The {\bf algebra of tensorial distinguished fields} $DT\left( {\cal E}%
\right) $ (d-fields, d-tensors, d-objects) on ${\cal E}$ is introduced as
the tensor algebra ${\cal T} =\{ {\cal T}_{qs}^{pr}\}$ of the v-bundle
${\cal E}_{\left( d\right) },$
 $p_d:{\cal HE\oplus VE\rightarrow E.\,\ }$ An
element ${\bf t}\in {\cal T}_{qs}^{pr},$ d-tensor field of type $\left(
\begin{array}{cc}
p & r \\
q & s
\end{array}
\right) ,$ can be written in local form as%
$$
{\bf t}=t_{j_1...j_qb_1...b_r}^{i_1...i_pa_1...a_r}\left( u\right) \delta
_{i_1}\otimes ...\otimes \delta _{i_p}\otimes \partial _{a_1}\otimes
...\otimes \partial _{a_r}\otimes d^{j_1}\otimes ...\otimes d^{j_q}\otimes
\delta ^{b_1}...\otimes \delta ^{b_r}.
$$

We shall respectively use denotations ${\cal X\left( E\right) }$ (or ${\cal %
X\ } {\left( M\right) ),\ } \Lambda ^p\left( {\cal E}\right) $ $\left(
\mbox{or }\Lambda ^p\left( M\right) \right) $ and ${\cal F\left( E\right) }$
(or ${\cal F}$ $\left( M\right) $) for the module of d-vector fields on $%
{\cal E}$ (or $M$ ), the exterior algebra of p-forms on ${\cal E\ }$(or $M)$
and the set of real functions on ${\cal E\ }$(or $M).$

In general, d-objects on ${\cal E\ }$ are introduced as geometric objects
with various group and coordinate transforms coordinated with the
N-connection structure on ${\cal E.\ }$ For example, a d-connection $D{\bf \
}$on ${\cal E\ }$ is defined as a linear connection $D$ on $E$ conserving
under a parallelism the global decomposition (2.3) into horizontal and
vertical subbundles of ${\cal TE}$ .

A N-connection in ${\cal E}$ induces a corresponding decomposition of
d-tensors into sums of horizontal and vertical parts, for example, for every
d-vector $X\in {\cal X\left( E\right) }$ and 1-form $\widetilde{X}\in
\Lambda ^1\left( {\cal E}\right) $ we have respectively
\f
X=hX+vX{\bf \ \quad }\mbox{and \quad }\widetilde{X}=h\widetilde{X}+v%
\widetilde{X}~. \n{2.7}
\e
In consequence, we can associate to every d-covariant derivation $D_X$ $%
=X\circ D$ two new operators of h- and v-covariant derivations defined
respectively as
$$
D_X^{(h)}Y=D_{hX}Y\quad \mbox{ and \quad }D_X^{\left( v\right) }Y=D_{vX}Y%
{\bf ,\quad }\forall Y{\bf \in }{\cal X\left( E\right) ,}
$$
for which the following conditions hold:%
\f
D_XY{\bf =}D_X^{(h)}Y{\bf \ +}D_X^{(v)}Y{\bf ,} \n{2.8}
\e
$$
D_X^{(h)}f=(hX{\bf )}f\mbox{ \quad and\quad }D_X^{(v)}f=(vX{\bf )}f,\quad X,Y%
{\bf \in }{\cal X\left( E\right) ,}f\in {\cal F}\left(
M\right) .
$$

We define a {\bf metric structure }${\bf G\ }$in the total space $E$ of
v-bundle ${\cal E=}$ $\left( E,p,M\right) $ over a connected and paracompact
base $M$ as a symmetric covariant tensor field of type $\left( 0,2\right) $,
$G_{\alpha \beta ,}$being nondegenerate and of constant signature on
$E.$

Nonlinear connection ${\bf N}$ and metric ${\bf G}$ structures on ${\cal E}$
are mutually compatible it there are satisfied the conditions:

\f
{\bf G}\left( \delta _i,\partial _a\right) =0,\mbox{or equivalently, }%
G_{ia}\left( u\right) -N_i^b\left( u\right) h_{ab}\left( u\right) =0, \n{2.9}
\e
where $h_{ab}={\bf G}\left( \partial _a,\partial _b\right) $ and $G_{ia}=%
{\bf G}\left( \partial _i,\partial _a\right) ,\,$ which gives%
\f
N_i^b\left( u\right) =h^{ab}\left( u\right) G_{ia}\left( u\right) \n{2.10}
\e
(the matrix $h^{ab}$ is inverse to $h_{ab}).$ In consequence one
obtains the following decomposition of metric :%
\f
{\bf G}(X,Y){\bf =hG}(X,Y)+{\bf vG}(X,Y){\bf ,} \n{2.11}
\e
where the d-tensor ${\bf hG}(X,Y){\bf =G}(hX,hY)$ is of type $\left(
\begin{array}{cc}
0 & 0 \\
2 & 0
\end{array}
\right) $ and the d-tensor ${\bf vG}(X,Y){\bf =G}(vX,vY)$ is of type $\left(
\begin{array}{cc}
0 & 0 \\
0 & 2
\end{array}
\right) .$ With respect to la-basis (2.5) the d-metric ( 2.11) is written as%
\f
{\bf G}=g_{\alpha \beta }\left( u\right) \delta ^\alpha \otimes \delta
^\beta =g_{ij}\left( u\right) d^i\otimes d^j+h_{ab}\left( u\right) \delta
^a\otimes \delta ^b, \n{2.12}
\e
where $g_{ij}={\bf G}\left( \delta _i,\delta _j\right) .$
\vskip30pt

A metric structure of type (2.11) (equivalently, of type (2.12)) or a metric
on $E$ with components satisfying constraints (2.9), equivalently (2.10))
defines an adapted to the given N-connection inner (d-scalar) product on the
tangent bundle ${\cal TE.}$

We shall say that a d-connection $\widehat{D}_X$ is compatible with the
d-scalar product on ${\cal TE\ }$ (i.e. is a standard d-connection) if
\vskip20pt
\f
\widehat{D}_X\left( {\bf X\cdot Y}\right) =\left( \widehat{D}_X{\bf Y}%
\right) \cdot {\bf Z+Y\cdot }\left( \widehat{D}_X{\bf Z}\right) ,\forall
{\bf X,Y,Z}{\bf \in }{\cal X\left( E\right) .} \n{2.13}
\e
An arbitrary d-connection $D_X$ differs from the standard one $\widehat{D}_X$
by an operator $\widehat{P}_X\left( u\right) =\{X^\alpha \widehat{P}_{\alpha
\beta }^\gamma \left( u\right) \},$ called the deformation d-tensor with
respect to $\widehat{D}_X,$ which is just a d-linear transform of
 ${\cal E}_u,$ $\forall u\in {\cal E.}$ The explicit form of $\widehat{P}_X$ can be
found by using the corresponding axiom defining linear connections [14]
$$
\left( D_X-\widehat{D}_X\right) fZ=f\left( D_X-\widehat{D}_X\right) Z{\bf ,}
$$
written with respect to la-bases (2.4) and (2.5). From the last expression
we obtain
\f
\widehat{P}_X\left( u\right) =\left[ (D_X-\widehat{D}_X)\delta _\alpha
\left( u\right) \right] \delta ^\alpha \left( u\right) , \n{2.14}
\e
therefore
\f
D_XZ{\bf \ }=\widehat{D}_XZ{\bf \ +}\widehat{P}_XZ  . \n{2.15}
\e
\vskip10pt

A d-connection $D_X$ is {\bf metric (}or {\bf compatible }with metric ${\bf G%
}$) on ${\cal E}$ if%
\f
D_X{\bf G} =0,\forall X{\bf \in }{\cal X\left( E\right) } . \n{2.16}
\e
Locally adapted components $\Gamma _{\beta \gamma }^\alpha $ of a
d-connection $D_\alpha =(\delta _\alpha \circ D)$ are defined by the
equations%
$$
D_\alpha \delta _\beta =\Gamma _{\alpha \beta }^\gamma \delta _\gamma ,
$$
from which one immediately follows%
\f
\Gamma _{\alpha \beta }^\gamma \left( u\right) =\left( D_\alpha \delta
_\beta \right) \circ \delta ^\gamma . \n{2.17}
\e
\vskip15pt
The operations of h- and v-covariant derivations, $D_k^{(h)}=%
\{L_{jk}^i,L_{bk\;}^a\}$ and $D_c^{(v)}=\{C_{jk}^i,C_{bc}^a\}$ (see (2.8)),
are introduced as corresponding h- and v-parametrizations of (2.17):%
\f
L_{jk}^i=\left( D_k\delta _j\right) \circ d^i,\quad L_{bk}^a=\left(
D_k\partial _b\right) \circ \delta ^a \n{2.18}
\e
and%
\f
C_{jc}^i=\left( D_c\delta _j\right) \circ d^i,\quad C_{bc}^a=\left(
D_c\partial _b\right) \circ \delta ^a. \n{2.19}
\e
A set of components (2.18) and (2.19), $D\Gamma =\left(
L_{jk}^i,L_{bk}^a,C_{jc}^i,C_{bc}^a\right) ,\,$ completely defines the local
action of a d-connection $D$ in ${\cal E.\,}$For instance, taken a d-tensor
field of type $\left(
\begin{array}{cc}
1 & 1 \\
1 & 1
\end{array}
\right) ,$ ${\bf t}=t_{jb}^{ia}\delta _i\otimes \partial _a\otimes \partial
^j\otimes \delta ^b,$ and a d-vector ${\bf X}=X^i\delta _i+X^a\partial _a$
we have%
$$
D_X{\bf t=}D_X^{(h)}{\bf t+}D_X^{(v)}{\bf t=}\left(
X^kt_{jb|k}^{ia}+X^ct_{jb\perp c}^{ia}\right) \delta _i\otimes \partial
_a\otimes d^j\otimes \delta ^b,
$$
where the h-covariant derivative is written as%
$$
t_{jb|k}^{ia}=\frac{\delta t_{jb}^{ia}}{\delta x^k}%
+L_{hk}^it_{jb}^{ha}+L_{ck}^at_{jb}^{ic}-L_{jk}^ht_{hb}^{ia}-L_{bk}^ct_{jc}^{ia}
$$
and the v-covariant derivative is written as%
$$
t_{jb\perp c}^{ia}=\frac{\partial t_{jb}^{ia}}{\partial y^c}%
+C_{hc}^it_{jb}^{ha}+C_{dc}^at_{jb}^{id}-C_{jc}^ht_{hb}^{ia}-C_{bc}^dt_{jd}^{ia}.
$$
For a scalar function $f\in {\cal F\left( E\right) }$ we have
$$
D_k^{(h)}=\frac{\delta f}{\delta x^k}=\frac{\partial f}{\partial x^k}-N_k^a
\frac{\partial f}{\partial y^a}\mbox{ and }D_c^{(v)}f=\frac{\partial f}{%
\partial y^c}.
$$
\vskip10pt
We emphasize that the geometry of connections in a v-bundle ${\cal E}$ is
very reach. If a triple of fundamental geometric objects $\left( N_i^a\left(
u\right) ,\Gamma _{\beta \gamma }^\alpha \left( u\right) ,G_{\alpha \beta
}\left( u\right) \right) $ is fixed on ${\cal E\ ,}$ really, a
multiconnection structure (with corresponding different rules of covariant
derivation, which are, or not, mutually compatible and with the same, or
not, induced d-scalar products in ${\cal TE)}$ is defined on this v-bundle.
For instance, we enumerate some of connections and covariant derivations
which can present interest in investigation of locally anisotropic
gravitational and matter field interactions:

\begin{enumerate}
\item  Every N-connection in ${\cal E,}$ with coefficients $N_i^a\left(
x,y\right) $ being differentiable on y-variables induces a structure of
linear connection $\widetilde{N}_{\beta \gamma }^\alpha ,$ where $\widetilde{%
N}_{bi}^a=\frac{\partial N_i^a}{\partial y^b}$ and $\widetilde{N}%
_{bc}^a\left( x,y\right) =0.$ For some $Y\left( u\right) =Y^i\left( u\right)
\partial _i+Y^a\left( u\right) \partial _a$ and $B\left( u\right) =B^a\left(
u\right) \partial _a$ one writes%
$$
D_Y^{(\widetilde{N})}B=\left[ Y^i\left( \frac{\partial B^a}{\partial x^i}+%
\widetilde{N}_{bi}^aB^b\right) +Y^b\frac{\partial B^a}{\partial y^b}\right]
\frac \partial {\partial y^a}.
$$

\item  The d-connection of Berwald type [15]%
\vskip20pt
\f
\Gamma _{\beta \gamma }^{(B)\alpha }=\left( L_{jk}^i,\frac{\partial N_k^a}{%
\partial y^b},0,C_{bc}^a\right) , \n{2.20}
\e
where
$$
L_{.jk}^i\left( x,y\right) =\frac 12g^{ir}\left( \frac{\delta g_{jk}}{\delta
x^k}+\frac{\delta g_{kr}}{\delta x^j}-\frac{\delta g_{jk}}{\delta x^r}%
\right) ,
$$
\vskip10pt
\f
C_{.bc}^a\left( x,y\right) =\frac 12h^{ad} \left( \frac{\partial h_{bd}}
{\partial y^c}+ \frac{\partial h_{cd}}{\partial y^b}-\frac{\partial h_{bc}}
{\partial y^d}\right) , \n{2.21}
\e
\vskip20pt
which is hv-metric, i.e. $D_k^{(B)}g_{ij}=0$ and $D_c^{(B)}h_{ab}=0.$
\vskip20pt
\item  The canonical d-connection ${\bf \Gamma ^{(c)}}$ associated to a
metric ${\bf G}$ of type (2.12) $\Gamma _{\beta \gamma }^{(c)\alpha }=\left(
L_{jk}^{(c)i},L_{bk}^{(c)a},C_{jc}^{(c)i},C_{bc}^{(c)a}\right) ,$ with
coefficients%
$$
L_{jk}^{(c)i}=L_{.jk}^i,C_{bc}^{(c)a}=C_{.bc}^a \mbox{ (see (2.21)}
$$
$$
L_{bi}^{(c)a}=\widetilde{N}_{bi}^a+\frac 12h^{ac}\left( \frac{\delta h_{bc}}{%
\delta x^i}-\widetilde{N}_{bi}^dh_{dc}-\widetilde{N}_{ci}^dh_{db}\right) ,
$$
\vskip15pt
\f
C_{jc}^{(c)i}=\frac 12g^{ik}\frac{\partial g_{jk}}{\partial y^c}. \n{2.22}
\e
\vskip15pt
This is a metric d-connection which satisfies conditions
$$
D_k^{(c)}g_{ij}=0,D_c^{(c)}g_{ij}=0,D_k^{(c)}h_{ab}=0,D_c^{(c)}h_{ab}=0.
$$

\item  We can consider N-adapted Christoffel d-symbols%
\f
\widetilde{\Gamma }_{\beta \gamma }^\alpha =\frac 12G^{\alpha \tau }\left(
\delta _\gamma G_{\tau \beta }+\delta _\beta G_{\tau \gamma }-\delta
G_{\beta \gamma }\right) , \n{2.23}
\e
which have the components of d-connection $\widetilde{\Gamma }_{\beta \gamma
}^\alpha =\left( L_{jk}^i,0,0,C_{bc}^a\right) ,$ with $L_{jk}^i$ and $%
C_{bc}^a$ as in (2.21) if $G_{\alpha \beta }$ is taken in the form
(2.12).
\end{enumerate}

Arbitrary linear connections on a v-bundle ${\cal E}$ can be also
characterized by theirs deformation tensors (see (2.15)) with respect, for
instance, to d-connection (2.23):%
$$
\Gamma _{\beta \gamma }^{(B)\alpha }=\widetilde{\Gamma }_{\beta \gamma
}^\alpha +P_{\beta \gamma }^{(B)\alpha },\Gamma _{\beta \gamma }^{(c)\alpha
}=\widetilde{\Gamma }_{\beta \gamma }^\alpha +P_{\beta \gamma }^{(c)\alpha }
$$
or, in general,%
$$
\Gamma _{\beta \gamma }^\alpha =\widetilde{\Gamma }_{\beta \gamma }^\alpha
+P_{\beta \gamma }^\alpha ,
$$
where $P_{\beta \gamma }^{(B)\alpha },P_{\beta \gamma }^{(c)\alpha }$ and $%
P_{\beta \gamma }^\alpha $ are respectively the deformation d-tensors of
d-connections (2.20),\ (2.22), or of a general one.
\vskip40pt
{\large \em 2.2. Torsions and curvatures of nonlinear}

{\large \em \qquad and distinguished connections}
\vskip15pt
The curvature ${\bf \Omega }$$\,$ of a nonlinear connection ${\bf N}$ in a
v-bundle ${\cal E}$ can be defined as the Nijenhuis tensor field $N_v\left(
X,Y\right) $ associated to ${\bf N\ }$ [2] :
$$
{\bf \Omega }=N_v={\bf \left[ vX,vY\right] +v\left[ X,Y\right] -v\left[
vX,Y\right] -v\left[ X,vY\right] ,X,Y}\in {\cal X\left( E\right) .}
$$
In local form one has%
$$
{\bf \Omega }=\frac 12\Omega _{ij}^ad^i\bigwedge d^j\otimes \partial _a,
$$
where%
\f
\Omega _{ij}^a=\frac{\partial N_i^a}{\partial x^j}-\frac{\partial N_j^a}{%
\partial x^i}+N_i^b\widetilde{N}_{bj}^a-N_j^b\widetilde{N}_{bi}^a. \n{2.24}
\e

The torsion ${\bf T}$ of a d-connection ${\bf D\ }$ in ${\cal E}$ is defined
by the equation%
\f
{\bf T\left( X,Y\right) =XY_{\circ }^{\circ }T\doteq }D_X{\bf Y-}D_Y{\bf X\
-\left[ X,Y\right] .} \n{2.25}
\e
One holds the following h- and v-decompositions%
\f
{\bf T\left( X,Y\right) =T\left( hX,hY\right) +T\left( hX,vY\right) +T\left(
vX,hY\right) +T\left( vX,vY\right) .} \n{2.26}
\e
We consider the projections :${\bf hT\left( X,Y\right) ,vT\left(
hX,hY\right) ,hT\left( hX,hY\right) ,...}$ and say that, for instance, ${\bf %
hT\left( hX,hY\right) }$ is the h(hh)-torsion of ${\bf D}$ , ${\bf vT\left(
hX,hY\right) \ }$ is the v(hh)-torsion of ${\bf D}$ and so on.

The torsion (2.25) is locally determined by five d-tensor fields, torsions,
defined as
$$
T_{jk}^i={\bf hT}\left( \delta _k,\delta _j\right) \cdot d^i,\quad T_{jk}^a=%
{\bf vT}\left( \delta _k,\delta _j\right) \cdot \delta ^a,
$$
$$
P_{jb}^i={\bf hT}\left( \partial _b,\delta _j\right) \cdot d^i,\quad
P_{jb}^a={\bf vT}\left( \partial _b,\delta _j\right) \cdot \delta ^a,
$$
\f
S_{bc}^a={\bf vT}\left( \partial _c,\partial _b\right) \cdot \delta ^a.
 \n{2.27} \e
Using formulas (2.4),(2.5),(2.24) and (2.25) we can computer [2] in explicit
form the components of torsions (2.26) for a d-connection of type (2.18) and
(2.19):
$$
T_{.jk}^i=T_{jk}^i=L_{jk}^i-L_{kj}^i,\quad T_{ja}^i=C_{.ja}^i,
T_{aj}^i=-C_{ja}^i,  T_{.ja}^i=0,T_{.bc}^a=S_{.bc}^a=C_{bc}^a-C_{cb}^a,
$$
\f
T_{.ij}^a=\frac{\delta N_i^a}{\delta x^j}-\frac{\delta N_j^a}{\delta x^i}%
,\quad T_{.bi}^a=P_{.bi}^a=\frac{\partial N_i^a}{\partial y^b}%
-L_{.bj}^a,\quad T_{.ib}^a=-P_{.bi}^a. \n{2.28}
\e

The curvature ${\bf R}$ of a d-connection in ${\cal E}$ is defined by the
equation
\f
{\bf R\left( X,Y\right) Z=XY_{\bullet }^{\bullet }R\bullet Z}=D_XD_Y{\bf Z}%
-D_YD_X{\bf Z-}D_{[X,Y]}{\bf Z.} \n{2.29}
\e
One holds the next properties for the h- and v-decompositions of curvature:%
$$
{\bf vR\left( X,Y\right) hZ=0,\ hR\left( X,Y\right) vZ=0,}
$$
\f
{\bf R\left( X,Y\right) Z=hR\left( X,Y\right) hZ+vR\left( X,Y\right) vZ.}%
\n{2.30}
\e
From (2.29) and the equation ${\bf R\left( X,Y\right) =-R\left( Y,X\right) }$
we get that the curvature of a d-connection ${\bf D}$ in ${\cal E}$ is
completely determined by the following six d-tensor fields:%
$$
R_{h.jk}^{.i}=d^i\cdot {\bf R}\left( \delta _k,\delta _j\right) \delta
_h,~R_{b.jk}^{.a}=\delta ^a\cdot {\bf R}\left( \delta _k,\delta _j\right)
\partial _b,
$$
$$
P_{j.kc}^{.i}=d^i\cdot {\bf R}\left( \partial _c,\partial _k\right) \delta
_j,~P_{b.kc}^{.a}=\delta ^a\cdot {\bf R}\left( \partial _c,\partial
_k\right) \partial _b,
$$
\f
S_{j.bc}^{.i}=d^i\cdot {\bf R}\left( \partial _c,\partial _b\right) \delta
_j,~S_{b.cd}^{.a}=\delta ^a\cdot {\bf R}\left( \partial _d,\partial
_c\right) \partial _b. \n{2.31}
\e
By a direct computation, using (2.4),(2.5),(2.18),(2.19) and (2.31) we get
[2]:
$$
R_{h.jk}^{.i}=\frac{\delta L_{.hj}^i}{\delta x^h}-\frac{\delta L_{.hk}^i}{%
\delta x^j}+L_{.hj}^mL_{mk}^i-L_{.hk}^mL_{mj}^i+C_{.ha}^iR_{.jk}^a,
$$
$$
R_{b.jk}^{.a}=\frac{\delta L_{.bj}^a}{\delta x^k}-\frac{\delta L_{.bk}^a}{%
\delta x^j}+L_{.bj}^cL_{.ck}^a-L_{.bk}^cL_{.cj}^a+C_{.bc}^aR_{.jk}^c,
$$
$$
P_{j.ka}^{.i}=\frac{\partial L_{.jk}^i}{\partial y^k}-\left( \frac{\partial
C_{.ja}^i}{\partial x^k}%
+L_{.lk}^iC_{.ja}^l-L_{.jk}^lC_{.la}^i-L_{.ak}^cC_{.jc}^i\right)
+C_{.jb}^iP_{.ka}^b,
$$
$$
P_{b.ka}^{.c}=\frac{\partial L_{.bk}^c}{\partial y^a}-\left( \frac{\partial
C_{.ba}^c}{\partial x^k}+L_{.dk}^{c%
\,}C_{.ba}^d-L_{.bk}^dC_{.da}^c-L_{.ak}^dC_{.bd}^c\right)
+C_{.bd}^cP_{.ka}^d,
$$
$$
S_{j.bc}^{.i}=\frac{\partial C_{.jb}^i}{\partial y^c}-\frac{\partial
C_{.jc}^i}{\partial y^b}+C_{.jb}^hC_{.hc}^i-C_{.jc}^hC_{hb}^i,
$$
\f
S_{b.cd}^{.a}=\frac{\partial C_{.bc}^a}{\partial y^d}-\frac{\partial
C_{.bd}^a}{\partial y^c}+C_{.bc}^eC_{.ed}^a-C_{.bd}^eC_{.ec}^a.
\n{2.32}
\e

Torsions (2.25) and curvatures (2.29) satisfy corresponding Ricci and
Bianchi identities (see [2]). We also note that torsions (2.28) and
curvatures (2.32) can be computed by particular cases of d-connections when
d-connections (2.20), (2.22) or (2.24) are used instead of (2.18) and (2.19)
.

The components of the Ricci d-tensor
$$
R_{\alpha \beta }=R_{\alpha .\beta \tau }^{.\tau }
$$
with respect to locally adapted frame (2.5) are as follows:%
$$
R_{ij}=R_{i.jk}^{.k},\quad R_{ia}=-^2P_{ia}=-P_{i.ka}^{.k},
$$
\f
R_{ai}=^1P_{ai}=P_{a.ib}^{.b},\quad R_{ab}=S_{a.bc}^{.c}.
\n{2.33}
\e
We point out that because, in general, $^1P_{ai}\neq ~^2P_{ia}$ the Ricci
d-tensor is non symmetric.

Having defined a d-metric of type (2.12) in ${\cal E}$ we can introduce the
scalar curvature of d-connection ${\bf D}$:
\f
{\overleftarrow{R}}=G^{\alpha \beta }R_{\alpha \beta }=R+S,{\n{2.34}}
\e
where $R=g^{ij}R_{ij}$ and $S=h^{ab}S_{ab}.$

For our further considerations it will be also useful to use an alternative
way of definition torsion (2.25) and curvature (2.29) by using the
commutator
\f
\Delta _{\alpha \beta }\doteq \nabla _\alpha \nabla _\beta -\nabla _\beta
\nabla _\alpha =2\nabla _{[\alpha }\nabla _{\beta ]}. \n{2.35}
\e
For components (2.28) of d-torsion we have
\f
\Delta _{\alpha \beta }f=T_{.\alpha \beta }^\gamma \nabla _\gamma f \n{2.36}
\e
for every scalar function $f\,\,$ on ${\cal E.}$ Curvature can be introduced
as an operator acting on arbitrary d-vector $V^\delta :$

\f
(\Delta _{\alpha \beta }-T_{.\alpha \beta }^\gamma \nabla _\gamma )V^\delta
=R_{~\gamma .\alpha \beta }^{.\delta }V^\gamma \n{2.37}
\e
(we note that in this work we shall follow conventions of Miron and
Anastasiei [2] on d-tensors; we can obtain corresponding Penrose and Rindler
abstract index formulas [12,13] just for a trivial N-connection structure
and by changing denotations for components of torsion and curvature in this
manner:\ $T_{.\alpha \beta }^\gamma \rightarrow T_{\alpha \beta }^{\quad
\gamma }$ and $R_{~\gamma .\alpha \beta }^{.\delta }\rightarrow R_{\alpha
\beta \gamma }^{\qquad \delta }).$

Here we also note that torsion and curvature of a d-connection on ${\cal E}$
satisfy generalized for la-spaces Ricci and Bianchi identities [2] which in
terms of components (2.36) and (2.37) are written respectively as%
\f
R_{~[\gamma .\alpha \beta ]}^{.\delta }+\nabla _{[\alpha }T_{.\beta \gamma
]}^\delta +T_{.[\alpha \beta }^\nu T_{.\gamma ]\nu }^\delta =0 \n{2.38}
\e
and%
\f
\nabla _{[\alpha }R_{|\nu |\beta \gamma ]}^{\cdot \sigma }+T_{\cdot [\alpha
\beta }^\delta R_{|\nu |.\gamma ]\delta }^{\cdot \sigma }=0. \n{2.39}
\e
Identities (2.38) and (2.39) can be proved similarly as in [12] by taking
into account that indices play a distinguished character.

We can also consider a la-generalization of the so-called conformal Weyl
tensor (see, for instance, [12]) which can be written as a d-tensor in this
form:%
$$
C_{\quad \alpha \beta }^{\gamma \delta }=R_{\quad \alpha \beta }^{\gamma
\delta }-\frac 4{n+m-2}R_{\quad [\alpha }^{[\gamma }~\delta _{\quad \beta
]}^{\delta ]}+$$
\f
\frac 2{(n+m-1)(n+m-2)}{\overleftarrow{R}~\delta _{\quad
[\alpha }^{[\gamma }~\delta _{\quad \beta ]}^{\delta ]}.} \n{2.40}
\e
This object is conformally invariant on la-spaces provided with d-connection
 generated by d-metric structures.
\vskip30pt
{\large \em 2.3. \ Field equations for la-gravity}
\vskip15pt
The Einstein equations and conservation laws on v-bundles provided with
N-connection structures are studied in detail in [2,16-19]. In [9] we proved
that the la-gravity can be formulated in a gauge like manner and analyzed
the conditions when the Einstein la-gravitational field equations are
equivalent to a corresponding form of Yang-Mills equations. In this
subsection we shall write the la-gravitational field equations in a form
more convenient for theirs equivalent reformulation in la-spinor variables.

We define d-tensor $\Phi _{\alpha \beta }$ as to satisfy conditions
\f
-2\Phi _{\alpha \beta }\doteq R_{\alpha \beta }-\frac 1{n+m}\overleftarrow{R}%
g_{\alpha \beta } \n{2.41}
\e
which is the torsionless part of the Ricci tensor for locally isotropic
spaces [12], i.e. $\Phi _\alpha ^{~~\alpha }\doteq 0$.\ The Einstein
equations on la-spaces
\f
\overleftarrow{G}_{\alpha \beta }+\lambda g_{\alpha \beta }=\kappa E_{\alpha
\beta }, \n{2.42}
\e
where%
\f
\overleftarrow{G}_{\alpha \beta }=R_{\alpha \beta }-\frac 12\overleftarrow{R}%
g_{\alpha \beta } \n{2.43}
\e
is the Einstein d-tensor, $\lambda $ and $\kappa $ are correspondingly the
cosmological and gravitational constants and by $E_{\alpha \beta }$
is denoted the locally anisotropic energy-momentum d-tensor [2], can
be rewritten in equivalent form:%
\f
\Phi _{\alpha \beta }=-\frac \kappa 2(E_{\alpha \beta }-\frac 1{n+m}E_\tau
^{~\tau }~g_{\alpha \beta }). \n{2.44}
\e

Because la-spaces generally have nonzero torsions we shall add to (2.44)
(equivalently to 2.42) a system of algebraic d-field equations with the
source $S_{~\beta \gamma }^\alpha $ being the locally anisotropic spin
density of matter (if we consider a variant of locally anisotropic
Einstein-Cartan theory):%
\f
T_{~\alpha \beta }^\gamma +2\delta _{~[\alpha }^\gamma T_{~\beta ]\delta
}^\delta =\kappa S_{~\alpha \beta .}^\gamma \n{2.45}
\e
From (2.38 ) and (2.45) one follows the conservation law of locally
anisotropic spin matter:%
$$
\nabla _\gamma S_{~\alpha \beta }^\gamma -T_{~\delta \gamma }^\delta
S_{~\alpha \beta }^\gamma =E_{\beta \alpha }-E_{\alpha \beta }.
$$

Finally, in this section, we remark that all presented geometric
constructions contain those elaborated for generalized Lagrange spaces [2]
(for which a tangent bundle $TM$ is considered instead of a v-bundle ${\cal E%
}$ ). We also note that the Lagrange (Finsler) geometry is characterized by
a metric of type (2.12) with components parametized as $g_{ij}=\frac 12%
\frac{\partial ^2{\cal L}}{\partial y^i\partial y^j}$ $\left( g_{ij}=\frac 12%
\frac{\partial ^2\Lambda ^2}{\partial y^i\partial y^j}\right) $ and $%
h_{ij}=g_{ij},$ where ${\cal L=L}$ $(x,y)$ $\left( \Lambda =\Lambda \left(
x,y\right) \right) $ is a Lagrangian $\left( \text{Finsler metric}\right) $
on $TM$ (see details in [1-4]).
\vskip40pt
{\large \bf 3. Spinor techniques for distinguished vector spaces}
\vskip15pt
The purpose of this section is to show how a corresponding abstract spinor
technique entailing notational and calculations advantages can be developed
for arbitrary splits of dimensions of a d-vector space
${\cal F}= h{\cal F} \oplus v{\cal F}$,
 where $\dim h{\cal F} = n$ and $\dim v{\cal F} = m.$ For convenience we
shall also present some necessary coordinate expressions.

The problem of a rigorous definition of spinors on la-spaces (la-spinors,
d-spinors) was posed and solved [7,8] in the framework of the formalism of
Clifford and spinor structures on v-bundles provided with compatible
nonlinear and distinguished connections and metric. We introduced d-spinors
as corresponding objects of the Clifford d-algebra ${\cal C}\left( {\cal F},
G\right)$, defined for a d-vector space ${\cal F}$ in a standard manner
(see, for instance, [20]) and proved that operations with
${\cal C}\left({\cal F},G\right) \ $ can be reduced to
calculations for ${\cal C}\left(h{\cal F},g\right) $ and ${\cal C} \left(
v{\cal F},h\right) ,$ which are usual Clifford algebras of respective dimensions 2$%
^n$ and 2$^m$ (if it is necessary we can use quadratic forms $g$ and $h$
correspondingly
induced on $h{\cal F}$ and $v{\cal F}$ by a metric ${\bf G}$ (2.12)). Considering the
orthogonal subgroup $O{\bf \left( G\right) }\subset GL{\bf \left( G\right) }$
defined by a metric ${\bf G}$ we can define the d-spinor norm and
parametrize d-spinors by ordered pairs of elements of Clifford algebras $%
{\cal C}\left(h{\cal F},g\right) $ and
${\cal C}\left( v{\cal F},h\right) .$ We emphasize
that the splitting of a Clifford d-algebra associated to a v-bundle ${\cal E}
$ is a straightforward consequence of the global decomposition (2.3) defining
a N-connection structure in ${\cal E.}$

In this section, as a rule, we shall omit proofs which in most cases are
mechanical but rather tedious. We can apply the methods developed in
[11-14] in a straightforward manner on h- and v-subbundles in order to verify
the correctness of affirmations.
\vskip30pt
{\large \em 3.1.\  Clifford d-algebra, d-spinors and d-twistors}
\vskip15pt
In order to relate the succeeding constructions with Clifford d-algebras [7,8]
we consider a la-frame decomposition of the metric (2.12):%
\f
G_{\alpha \beta }\left( u\right) =l_\alpha ^{\widehat{\alpha }}\left(
u\right) l_\beta ^{\widehat{\beta }}\left( u\right) G_{\widehat{\alpha }
\widehat{\beta }}, \n{3.1}
\e
where the frame d-vectors and constant metric matrices are distinguished as

\f
l_\alpha ^{\widehat{\alpha }}\left( u\right) =\left(
\begin{array}{cc}
l_j^{\widehat{j}}\left( u\right) & 0 \\
0 & l_a^{\widehat{a}}\left( u\right)
\end{array}
\right) ,G_{\widehat{\alpha }\widehat{\beta }}\left(
\begin{array}{cc}
g_{\widehat{i}\widehat{j}} & 0 \\
0 & h_{\widehat{a}\widehat{b}}
\end{array}
\right) , \n{3.2}
\e
$g_{\widehat{i}\widehat{j}}$ and $h_{\widehat{a}\widehat{b}}$ are diagonal
matrices with $g_{\widehat{i}\widehat{i}}=$ $h_{\widehat{a}\widehat{a}}=\pm
1. $

To generate Clifford d-algebras we start with matrix equations%
\f
\sigma _{\widehat{\alpha }}\sigma _{\widehat{\beta }}+\sigma _{\widehat{%
\beta }}\sigma _{\widehat{\alpha }}=-G_{\widehat{\alpha }\widehat{\beta }%
}I, \n{3.3}
\e
where $I$ is the identity matrix, matrices $\sigma _{\widehat{\alpha }%
}\,(\sigma $-objects) act on a d-vector space ${\cal F} =
h{\cal F} \oplus v{\cal F}$ and theirs components are distinguished as
\f
\sigma _{\widehat{\alpha }}\,=\left\{ (\sigma _{\widehat{\alpha }})_{%
\underline{\beta }}^{\cdot \underline{\gamma }}=\left(
\begin{array}{cc}
(\sigma _{\widehat{i}})_{\underline{j}}^{\cdot \underline{k}} & 0 \\
0 & (\sigma _{\widehat{a}})_{\underline{b}}^{\cdot \underline{c}}
\end{array}
\right) \right\} , \n{3.4}
\e
indices \underline{$\beta $},\underline{$\gamma $},... refer to spin spaces
of type ${\cal S} = S_{(h)}\oplus S_{(v)}$ and underlined Latin indices
\underline{$j$},$\underline{k},...$ and $\underline{b},\underline{c},...$
refer respectively to a h-spin space ${\cal S}_{(h)}$ and a v-spin space $%
{\cal S}_{(v)},\ $which are correspondingly associated to a h- and
v-decomposition of a v-bundle ${\cal E}_{(d)} .$ The irreducible algebra of
matrices $\sigma _{\widehat{\alpha }}$ of minimal dimension $N\times N,$
where $N=N_{(n)}+N_{(m)},$ $\dim {\cal S}_{(h)}$=$N_{(n)}$ and
$\dim {\cal S}_{(v)}$=$N_{(m)},$ has these dimensions%
$$
{N_{(n)}= \left\{\begin{array}{rl}{ 2^{(n-1)/2},}& n=2k+1 \\
           {2^{n/2},\ } & n=2k; \end{array} \right.} \qquad \mbox{ and} \qquad
{N_{(m)}= \left\{\begin{array}{rl}{2^{(m-1)/2},}& m=2k+1 \\
           {2^{m/2},\ } & m=2k, \end{array} \right .}
$$
where $k=1,2,...$ .

The Clifford d-algebra is generated by sums on $n+1$ elements of form%
\f
A_1I+B^{\widehat{i}}\sigma _{\widehat{i}}+C^{\widehat{i}\widehat{j}}\sigma _{%
\widehat{i}\widehat{j}}+D^{\widehat{i}\widehat{j}\widehat{k}}\sigma _{%
\widehat{i}\widehat{j}\widehat{k}}+... \n{3.5}
\e
and sums of $m+1$ elements of form%
$$
A_2I+B^{\widehat{a}}\sigma _{\widehat{a}}+C^{\widehat{a}\widehat{b}}\sigma _{%
\widehat{a}\widehat{b}}+D^{\widehat{a}\widehat{b}\widehat{c}}\sigma _{%
\widehat{a}\widehat{b}\widehat{c}}+...
$$
with antisymmetric coefficients $C^{\widehat{i}\widehat{j}}=C^{[\widehat{i}
\widehat{j}]},C^{\widehat{a}\widehat{b}}=C^{[\widehat{a}\widehat{b]}},D^{%
\widehat{i}\widehat{j}\widehat{k}}=D^{[\widehat{i}\widehat{j}\widehat{k}%
]},D^{\widehat{a}\widehat{b}\widehat{c}}=D^{[\widehat{a}\widehat{b}\widehat{c%
}]},...$ and matrices $\sigma _{\widehat{i}\widehat{j}}=\sigma _{[\widehat{i}%
}\sigma _{\widehat{j}]},\sigma _{\widehat{a}\widehat{b}}=\sigma _{[\widehat{a%
}}\sigma _{\widehat{b}]},\sigma _{\widehat{i}\widehat{j}\widehat{k}}=\sigma
_{[\widehat{i}}\sigma _{\widehat{j}}\sigma _{\widehat{k}]},...$ . Really, we
have 2$^{n+1}$ coefficients $\left( A_1,C^{\widehat{i}\widehat{j}},D^{%
\widehat{i}\widehat{j}\widehat{k}},...\right) $ and 2$^{m+1}$ coefficients $%
\left( A_2,C^{\widehat{a}\widehat{b}},D^{\widehat{a}\widehat{b}\widehat{c}%
},...\right) $ of the Clifford algebra on ${\cal F.}$

For simplicity, in this subsection, we shall present the necessary geometric
constructions only for h-spin spaces ${\cal S}_{(h)}$ of dimension $N_{(n)}.$
Considerations for a v-spin space ${\cal S}_{(v)}$ are similar but
with proper characteristics for a dimension $N_{(m)}.$

In order to define the scalar (spinor) product on ${\cal S}_{(h)}$ we
introduce into consideration this finite sum (because of a finite number of
elements $\sigma _{[\widehat{i}\widehat{j}...\widehat{k}]}$ ):%
$$
^{(\pm )}E_{\underline{k}\underline{m}}^{\underline{i}\underline{j}}=\delta
_{\underline{k}}^{\underline{i}}\delta _{\underline{m}}^{\underline{j}%
}+\frac 2{1!}(\sigma _{\widehat{i}})_{\underline{k}}^{.\underline{i}}(\sigma
^{\widehat{i}})_{\underline{m}}^{.\underline{j}}+\frac{2^2}{2!}(\sigma _{%
\widehat{i}\widehat{j}})_{\underline{k}}^{.\underline{i}}(\sigma ^{\widehat{i%
}\widehat{j}})_{\underline{m}}^{.\underline{j}}+\frac{2^3}{3!}(\sigma _{%
\widehat{i}\widehat{j}\widehat{k}})_{\underline{k}}^{.\underline{i}}(\sigma
^{\widehat{i}\widehat{j}\widehat{k}})_{\underline{m}}^{.\underline{j}}+...
$$
which can be factorized as
\f
^{(\pm )}E_{\underline{k}\underline{m}}^{\underline{i}\underline{j}}=N_{(n)}
{ }^{(\pm )}\epsilon _{\underline{k}\underline{m}}{ }^{(\pm
)}\epsilon ^{\underline{i}\underline{j}}\mbox{ for }n=2k \n{3.6}
\e
and%
$$
^{(+)}E_{\underline{k}\underline{m}}^{\underline{i}\underline{j}}=2N_{(n)}
\epsilon _{\underline{k}\underline{m}}\epsilon ^{\underline{i}
\underline{j}},{ }^{(-)}E_{\underline{k}\underline{m}}^{\underline{i}
\underline{j}}=0 \mbox{ for }n=3(mod4),
$$
$$
^{(+)}E_{\underline{k}\underline{m}}^{\underline{i}\underline{j}}=0,
{ }^{(-)}E_{\underline{k}\underline{m}}^{\underline{i}\underline{j}}=2N_{(n)}
\epsilon _{\underline{k}\underline{m}}\epsilon ^{\underline{i}
\underline{j}} \mbox{ for }n=1(mod4).
$$

Antisymmetry of $\sigma _{\widehat{i}\widehat{j}\widehat{k}...}$ and the
construction of the objects (3.5) and (3.6) define the properties of $%
\epsilon $-objects $^{(\pm )}\epsilon _{\underline{k}\underline{m}}$ and $%
\epsilon _{\underline{k}\underline{m}}$ which have an eight-fold periodicity
on $n$ (see details in [13] and, with respect to la-spaces, [7]).

For even values of $n$ it is possible the decomposition of every h-spin
space ${\cal S}_{(h)}$into irreducible h-spin spaces ${\bf S}_{(h)}$
and ${\bf S}_{(h)}^{\prime }$ (one considers splitting
of h-indices, for instance, \underline{$l$}$=L\oplus L^{\prime },\underline{m%
}=M\oplus M^{\prime },... ;$ for v-indices we shall write $\underline{a}%
=A\oplus A^{\prime },\underline{b}=B\oplus B^{\prime },...)$ and defines
new $\epsilon $-objects
\f
\epsilon ^{\underline{l}\underline{m}}=\frac 12\left( ^{(+)}\epsilon ^{%
\underline{l}\underline{m}}+^{(-)}\epsilon ^{\underline{l}\underline{m}%
}\right) \mbox{ and }\widetilde{\epsilon }^{\underline{l}\underline{m}%
}=\frac 12\left( ^{(+)}\epsilon ^{\underline{l}\underline{m}}-^{(-)}\epsilon
^{\underline{l}\underline{m}}\right) \n{3.7}
\e
We shall omit similar formulas for $\epsilon $-objects with lower indices.

We can verify, by using expressions (3.6) and straightforward calculations,
these parametrizations on symmetry properties of $\epsilon $-objects (3.7)
$$
\epsilon ^{\underline{l}\underline{m}}=\left(
\begin{array}{cc}
\epsilon ^{LM}=\epsilon ^{ML} & 0 \\
0 & 0
\end{array}
\right) \mbox{ and }
\widetilde{\epsilon }^{\underline{l}\underline{m}}=\left(
\begin{array}{cc}
0 & 0 \\
0 & \widetilde{\epsilon }^{LM}=\widetilde{\epsilon }^{ML}
\end{array}
\right) \mbox{ for }n=0(mod8);
$$
$$
\epsilon ^{\underline{l}\underline{m}} =
-\frac 12 {}^{(-)}\epsilon ^{\underline{l}\underline{m}} =
\epsilon ^{\underline{m}\underline{l}},\mbox{ where }%
^{(+)}\epsilon ^{\underline{l}\underline{m}}=0,\mbox{ and }$$  $$\widetilde{%
\epsilon }^{\underline{l}\underline{m}}=-\frac 12 {}^{(-)}\epsilon ^{\underline{%
l}\underline{m}}=\widetilde{\epsilon }^{\underline{m}\underline{l}}\mbox{
for }n=1(mod8);
$$
$$
\epsilon ^{\underline{l}\underline{m}}=\left(
\begin{array}{cc}
0 & 0 \\
\epsilon ^{L^{\prime }M} & 0
\end{array}
\right) \mbox{ and } \widetilde{\epsilon }^{\underline{l}\underline{m}%
}=\left(
\begin{array}{cc}
0 & \widetilde{\epsilon }^{LM^{\prime }}=-\epsilon ^{M^{\prime }L} \\ 0 & 0
\end{array}
\right) \mbox{ for }n=2(mod8);
$$
$$
\epsilon ^{\underline{l}\underline{m}}=-\frac 12 {}^{(+)}\epsilon ^{\underline{l%
}\underline{m}}=-\epsilon ^{\underline{m}\underline{l}},\mbox{ where }%
^{(-)}\epsilon ^{\underline{l}\underline{m}}=0,\mbox{ and }$$
$$\widetilde{%
\epsilon }^{\underline{l}\underline{m}}=\frac 12 {}^{(+)}\epsilon ^{\underline{l%
}\underline{m}}=-\widetilde{\epsilon }^{\underline{m}\underline{l}}\mbox{
for }n=3(mod8);
$$
$$
\epsilon ^{\underline{l}\underline{m}}=\left(
\begin{array}{cc}
\epsilon ^{LM}=-\epsilon ^{ML} & 0 \\
0 & 0
\end{array}
\right) \mbox{ and }  \widetilde{\epsilon }^{\underline{l}\underline{m}%
}=\left(
\begin{array}{cc}
0 & 0 \\
0 & \widetilde{\epsilon }^{LM}=-\widetilde{\epsilon }^{ML}
\end{array}
\right) \mbox{ for }n=4(mod8);
$$
$$
\epsilon ^{\underline{l}\underline{m}}=-\frac 12 {}^{(-)}\epsilon ^{\underline{l%
}\underline{m}}=-\epsilon ^{\underline{m}\underline{l}},\mbox{ where }%
^{(+)}\epsilon ^{\underline{l}\underline{m}}=0,\mbox{ and }$$
$$\widetilde{%
\epsilon }^{\underline{l}\underline{m}}=-\frac 12
{}^{(-)}\epsilon ^{\underline{%
l}\underline{m}} =-\widetilde{\epsilon }^{\underline{m}\underline{l}}\mbox{
for }n=5(mod8);
$$
$$
\epsilon ^{\underline{l}\underline{m}}=\left(
\begin{array}{cc}
0 & 0 \\
\epsilon ^{L^{\prime }M} & 0
\end{array}
\right) \mbox{ and } \widetilde{\epsilon }^{\underline{l}\underline{m}%
}=\left(
\begin{array}{cc}
0 & \widetilde{\epsilon }^{LM^{\prime }}=\epsilon ^{M^{\prime }L} \\ 0 & 0
\end{array}
\right) \mbox{ for }n=6(mod8);
$$
$$
\epsilon ^{\underline{l}\underline{m}}=\frac 12 {}^{(-)}\epsilon ^{\underline{l}%
\underline{m}}=\epsilon ^{\underline{m}\underline{l}},\mbox{ where }%
{}^{(+)}\epsilon ^{\underline{l}\underline{m}}=0,\mbox{ and } $$
\f  \widetilde{%
\epsilon }^{\underline{l}\underline{m}}=-\frac 12 {}^{(-)}\epsilon ^{\underline{%
l}\underline{m}}=\widetilde{\epsilon }^{\underline{m}\underline{l}}\mbox{
for }n=7(mod8). \n{3.8}
\e

Let denote reduced and irreducible h-spinor spaces in a form pointing to the
symmetry of spinor inner products in dependence of values $n=8k+l$ ($%
k=01,2,...;l=1,2,...7)$ of the dimension of the horizontal subbundle (we
shall write respectively $\bigtriangleup $ and $\circ $ for antisymmetric
and symmetric inner products of reduced spinors and $\diamondsuit
=(\bigtriangleup ,\circ )$ and $\widetilde{\diamondsuit }=(\circ
,\bigtriangleup )$ for corresponding parametrizations of inner products, in
brief {\it i.p.}, of irreducible spinors; properties of scalar products of
spinors are defined by $\epsilon $-objects (3.8); we shall use $\Diamond $
for a general {\it i.p.} when the symmetry is not pointed out):%
$$
{\cal S}_{(h)}{ }\left( 8k\right) ={\bf S_{\circ }\oplus S_{\circ
}^{\prime };\quad }$$ $${\cal S}_{(h)}{ }\left( 8k+1\right) ={\cal S}_{\circ
}^{(-)}\ \mbox{({\it i.p.} is defined by an }^{(-)}\epsilon \mbox{-object);}
$$
$$
{\cal S}_{(h)} { }\left( 8k+2\right) =\{
\begin{array}{c}
{\cal S}_{\Diamond }=({\bf S}_{\Diamond },{\bf S}_{\Diamond }),\mbox{ or}
\\ {\cal S}_{\Diamond }^{\prime }=({\bf S}_{\widetilde{\Diamond }%
}^{\prime }, {\bf S}_{\widetilde{\Diamond }}^{\prime });
\end{array}
\qquad$$ $$ {\cal S}_{(h)} \left( 8k+3\right) ={\cal S}_{\bigtriangleup
}^{(+)}\ \mbox{({\it i.p.} is defined by an }^{(+)}\epsilon \mbox{-object);}
$$
$$
{\cal S}_{(h)} \left( 8k+4\right) ={\bf S}_{\bigtriangleup }\oplus
{\bf S}_{\bigtriangleup }^{\prime };\quad $$  $${\cal S}_{(h)} \left(
8k+5\right) ={\cal S}_{\bigtriangleup }^{(-)}\ \mbox{({\it i.p. }is defined
by an }^{(-)}\epsilon \mbox{-object),}
$$
$$
{\cal S}_{(h)} \left( 8k+6\right) =\{
\begin{array}{c}
{\cal S}_{\Diamond } = ({\bf S}_{\Diamond },{\bf S}_{\Diamond }),\mbox{ or}
\\ {\cal S}_{\Diamond }^{\prime }=({\bf S}_{\widetilde{\Diamond }%
}^{\prime },{\bf S}_{\widetilde{\Diamond }}^{\prime });
\end{array}  $$  \qquad
\f {\cal S}_{(h)} \left( 8k+7\right) ={\cal S}_{\circ }^{(+)}\ %
\mbox{({\it i.p. } is defined by an }^{(+)}\epsilon \mbox{-object)}. \n{3.9}
\e
We note that by using corresponding $\epsilon $-objects we can lower and
rise indices of reduced and irreducible spinors (for $n=2,6(mod4)$ we can
exclude primed indices, or inversely, see details in [11-13,7]).

The similar v-spinor spaces are denoted by the same symbols as in (3.9)
provided with a left lower mark ''$|"$ and parametrized with respect to the
values $m=8k^{\prime }+l$ (k'=0,1,...; l=1,2,...,7) of the dimension of the
vertical subbundle, for example, as
\f
{\cal S}_{(v)} ( 8k^{\prime })={\bf S}_{|\circ }\oplus {\bf S}_{|\circ
}^{\prime },{\cal S}_{(v)} \left( 8k+1\right) ={\cal S}_{|\circ
}^{(-)},... \n{3.10}
\e
We use '' $\widetilde {}$ ''-overlined symbols,
\f
{\widetilde {\cal S}}_{(h)}\left( 8k\right) ={\widetilde{\bf S}}_{\circ
}\oplus \widetilde{S}_{\circ }^{\prime },{\widetilde {\cal S}}_{(h)}\left(
8k+1\right) ={\widetilde{\cal S}}_{\circ }^{(-)},... \n{3.11}
\e
and
\f
{\widetilde {\cal S}}_{(v)} ( 8k^{\prime })={\widetilde{\bf S}}_{|\circ
}\oplus {\widetilde S}_{|\circ }^{\prime },{\widetilde {\cal S}}_{(v)} \left(
8k^{\prime }+1\right) ={\widetilde {\cal S}}_{|\circ }^{(-)},... \n{3.12}
\e
respectively for the dual to (3.9) and (3.10) spinor spaces.

The spinor spaces (3.9)-(3.12) are called the prime spinor spaces, in brief
p-spinors. They are considered as building blocks of distinguished
(n,m)-spinor spaces constructed in this manner:%
$$
{\cal S}(_{\circ \circ ,\circ \circ })={\bf S_{\circ }\oplus S_{\circ
}^{\prime }\oplus S_{|\circ }\oplus S_{|\circ }^{\prime },}{\cal S}(_{\circ
\circ ,\circ }\mid ^{\circ })={\bf S_{\circ }\oplus S_{\circ }^{\prime
}\oplus S_{|\circ }\oplus \widetilde{S}_{|\circ }^{\prime },}
$$
$$
{\cal S}(_{\circ \circ ,}\mid ^{\circ \circ })={\bf S_{\circ }\oplus
S_{\circ }^{\prime }\oplus \widetilde{S}_{|\circ }\oplus \widetilde{S}%
_{|\circ }^{\prime },}{\cal S}(_{\circ }\mid ^{\circ \circ \circ })={\bf %
S_{\circ }\oplus \widetilde{S}_{\circ }^{\prime }\oplus \widetilde{S}%
_{|\circ }\oplus \widetilde{S}_{|\circ }^{\prime },}
$$
$$
...............................................
$$
$$
{\cal S}(_{\triangle }, _{\triangle })={\cal S}_{\triangle
}^{(+)}\oplus S_{|\bigtriangleup }^{(+)},S ( _{\triangle
},^{\triangle })={\cal S}_{\triangle }^{(+)}\oplus \widetilde{S}_{|\triangle
}^{(+)},
$$
$$
................................
$$
\f
{\cal S}(_{\triangle }|^{\circ },_\diamondsuit )={\bf S}_{\triangle
}\oplus \widetilde{S_{\circ }}^{\prime }\oplus {\cal S}_{|\diamondsuit },%
{\cal S}(_{\triangle }|^{\circ },^\diamondsuit )={\bf S}_{\triangle
}\oplus \widetilde{S_{\circ }}^{\prime }\oplus {\cal \widetilde{S}}%
_{|}^\diamondsuit , \n{3.13}
\e
$$
................................
$$
Considering the operation of dualization of prime components in (3.13) we
can generate different isomorphic variants of distinguished (n,m)-spinor
spaces.

We define a d-spinor space ${\cal S}_{(n,m)}\ $ as a direct sum of a
horizontal and a vertical spinor spaces of type (3.9), for instance,
$$
{\cal S}_{(8k,8k^{\prime })} = {\bf S}_{\circ }\oplus {\bf S}_{\circ }^{\prime
}\oplus {\bf S}_{|\circ }\oplus {\bf S}_{|\circ }^{\prime },
 {\cal S}_{(8k,8k^{\prime}+1)}\ =
{\bf S}_{\circ }\oplus{\bf S}_{\circ }^{\prime }\oplus {\cal S}_{|\circ
}^{(-)}, ...,$$  $$ {\cal S}_{(8k+4,8k^{\prime }+5)} = {\bf S}_{\triangle
}\oplus {\bf S}_{\triangle }^{\prime }\oplus {\cal S}_{|\triangle }^{(-)},...
$$
The scalar product on a ${\cal S}_{(n,m)}\ $ is induced by (corresponding to
fixed values of $n$ and $m$ ) $\epsilon $-objects (3.8) considered for h-
and v-components.

Having introduced d-spinors for dimensions $\left( n,m\right) $ we can
write out the generalization for la-spaces of twistor equations [13] by
using the distinguished $\sigma $-objects (3.4):%
\f
(\sigma _{(\widehat{\alpha }})_{|\underline{\beta }|}^{..\underline{\gamma }}%
\quad \frac{\delta \omega ^{\underline{\beta }}}{\delta u^{\widehat{\beta })}}%
=\frac 1{n+m} \quad G_{\widehat{\alpha }\widehat{\beta }}(\sigma ^{\widehat{%
\epsilon }})_{\underline{\beta }}^{..\underline{\gamma }} \quad
\frac{\delta \omega^{\underline{\beta }}}
{\delta u^{\widehat{\epsilon }}}, \n{3.14}
\e
where $\left| \underline{\beta }\right| $ denotes that we do not consider
symmetrization on this index. The general solution of (3.14) on the d-vector
space ${\cal F}$ looks like as
\f
\omega ^{\underline{\beta }}=\Omega ^{\underline{\beta }}+u^{\widehat{\alpha
}}(\sigma _{\widehat{\alpha }})_{\underline{\epsilon }}^{..\underline{\beta }%
}\Pi ^{\underline{\epsilon }}, \n{3.15}
\e
where $\Omega ^{\underline{\beta }}$ and $\Pi ^{\underline{\epsilon }}$ are
constant d-spinors. For fixed values of dimensions $n$ and $m$ we mast
analyze the reduced and irreducible components of h- and v-parts of
equations (3.14) and their solutions (3.15) in order to find the symmetry
properties of a d-twistor ${\bf Z^\alpha \ }$ defined as a pair of d-spinors%
$$
{\bf Z}^\alpha = (\omega ^{\underline{\alpha }},\pi _{\underline{%
\beta }}^{\prime }),
$$
where $\pi _{\underline{\beta }^{\prime }}=\pi _{\underline{\beta }^{\prime
}}^{(0)}\in {\widetilde{\cal S}}_{(n,m)}$ is a constant dual d-spinor. The
problem of definition of spinors and twistors on la-spaces was firstly
considered in [21] (see also [22,23]) in connection with the possibility to
extend the equations (3.14) and theirs solutions (3.15), by using nearly
autoparallel maps, on curved, locally isotropic or anisotropic, spaces.
\vskip30pt
{\large \em 3.2. Mutual transforms of d-tensors and d-spinors}
\vskip15pt
The spinor algebra for spaces of higher dimensions can not be considered as
a real alternative to the tensor algebra as for locally isotropic spaces of
dimensions $n=3,4$ [11-13]. The same holds true for la-spaces and we
emphasize that it is not quite convenient to perform a spinor calculus for
dimensions $n,m>>4$. Nevertheless, the concept of spinors is important for
every type of spaces, for instance, we can deeply understand the fundamental
properties of geometical objects on la-spaces, and we shall consider in this
subsection some questions concerning transforms of d-tensor objects into
d-spinor ones.
\vskip24pt
{\sf 3.2.1. Transformation of d-tensors into d-spinors}
\vskip12pt
In order to pass from d-tensors to d-spinors we must use $\sigma $-objects
(3.4) written in reduced or irreduced form  \quad (in dependence of fixed values of
dimensions $n$ and $m$ ):

\f
(\sigma _{\widehat{\alpha }})_{\underline{\beta }}^{\cdot \underline{\gamma }%
},~(\sigma ^{\widehat{\alpha }})^{\underline{\beta }\underline{\gamma }%
},~(\sigma ^{\widehat{\alpha }})_{\underline{\beta }\underline{\gamma }%
},...,(\sigma _{\widehat{a}})^{\underline{b}\underline{c}},...,(\sigma _{%
\widehat{i}})_{\underline{j}\underline{k}},...,(\sigma _{\widehat{a}%
})^{AA^{\prime }},...,(\sigma ^{\widehat{i}})_{II^{\prime }},....\n{3.16}
\e
It is obvious that contracting with corresponding $\sigma $-objects (3.16)
we can introduce instead of d-tensors indices the d-spinor ones, for
instance,%
$$
\omega ^{\underline{\beta }\underline{\gamma }}=(\sigma ^{\widehat{\alpha }%
})^{\underline{\beta }\underline{\gamma }}\omega _{\widehat{\alpha }},\quad
\omega _{AB^{\prime }}=(\sigma ^{\widehat{a}})_{AB^{\prime }}\omega _{%
\widehat{a}},\quad ...,\zeta _{\cdot \underline{j}}^{\underline{i}}=(\sigma
^{\widehat{k}})_{\cdot \underline{j}}^{\underline{i}}\zeta _{\widehat{k}%
},....
$$
For d-tensors containing groups of antisymmetric indices there is a more
simple procedure of theirs transforming into d-spinors because the objects
\f
(\sigma _{\widehat{\alpha }\widehat{\beta }...\widehat{\gamma }})^{%
\underline{\delta }\underline{\nu }},\quad (\sigma ^{\widehat{a}\widehat{b}%
...\widehat{c}})^{\underline{d}\underline{e}},\quad ...,(\sigma ^{\widehat{i}%
\widehat{j}...\widehat{k}})_{II^{\prime }},\quad ... \n{3.17}
\e
can be used for sets of such indices into pairs of d-spinor indices. Let us
enumerate some properties of $\sigma $-objects of type (3.17) (for
simplicity we consider only h-components having q indices $\widehat{i},%
\widehat{j},\widehat{k},...$ taking values from 1 to $n;$ the properties of
v-components can be written in a similar manner with respect to indices $%
\widehat{a},\widehat{b},\widehat{c}...$ taking values from 1 to $m$):%
\f
(\sigma _{\widehat{i}...\widehat{j}})^{\underline{k}\underline{l}}\mbox{
 is\ }\left\{ \
\begin{array}{c}
\mbox{symmetric on }\underline{k},\underline{l}\mbox{ for }n-2q\equiv
1,7~(mod~8); \\ \mbox{antisymmetric on }\underline{k},\underline{l}\mbox{
for }n-2q\equiv 3,5~(mod~8)
\end{array}
\right\} \n{3.18}
\e
for odd values of $n,$ and an object
$$
(\sigma _{\widehat{i}...\widehat{j}})^{IJ}~\left( (\sigma _{\widehat{i}...%
\widehat{j}})^{I^{\prime }J^{\prime }}\right)$$
\f  \mbox{ is\ }\left\{
\begin{array}{c}
\mbox{symmetric on }I,J~(I^{\prime },J^{\prime })\mbox{ for }n-2q\equiv
0~(mod~8); \\ \mbox{antisymmetric on }I,J~(I^{\prime },J^{\prime })\mbox{
for }n-2q\equiv 4~(mod~8)
\end{array}
\right\} \n{3.19}
\e
or%
\f
(\sigma _{\widehat{i}...\widehat{j}})^{IJ^{\prime }}=\pm (\sigma _{\widehat{i%
}...\widehat{j}})^{J^{\prime }I}\{
\begin{array}{c}
n+2q\equiv 6(mod8); \\
n+2q\equiv 2(mod8),
\end{array}
\n{3.20}
\e
with vanishing of the rest of reduced components of the d-tensor $(\sigma _{%
\widehat{i}...\widehat{j}})^{\underline{k}\underline{l}}$ with prime/unprime
sets of indices.
\vskip24pt
{\sf 3.2.2. Transformation of d-spinors into d-tensors; fundamental
d-spinors}
\vskip12pt
We can transform every d-spinor $\xi ^{\underline{\alpha }}=\left( \xi ^{%
\underline{i}},\xi ^{\underline{a}}\right) $ into a corresponding d-tensor.
For simplicity, we consider this construction only for a h-component $\xi ^{%
\underline{i}}$ on a h-space being of dimension $n$. The values%
\f
\xi ^{\underline{\alpha }}\xi ^{\underline{\beta }}(\sigma ^{\widehat{i}...%
\widehat{j}})_{\underline{\alpha }\underline{\beta }}\quad \left( n\mbox{ is
odd}\right) \n{3.21}
\e
or
\f
\xi ^I\xi ^J(\sigma ^{\widehat{i}...\widehat{j}})_{IJ}~\left( \mbox{or }\xi
^{I^{\prime }}\xi ^{J^{\prime }}(\sigma ^{\widehat{i}...\widehat{j}%
})_{I^{\prime }J^{\prime }}\right) ~\left( n\mbox{ is even}\right) \n{3.22}
\e
with a different number of indices $\widehat{i}...\widehat{j},$ taken
together,defines the h-spinor $\xi ^{\underline{i}}\,$ to an accuracy to the
sign. We emphasize that it is necessary to choose only those h-components
of d-tensors (3.21) (or (3.22)) which are symmetric on pairs of indices $%
\underline{\alpha }\underline{\beta }$ (or $IJ\,$ (or $I^{\prime }J^{\prime }
$ )) and the number $q$ of indices $\widehat{i}...\widehat{j}$ satisfies the
condition (as a respective consequence of the properties (3.18) and/or
(3.19), (3.20))%
\f
n-2q\equiv 0,1,7~(mod~8). \n{3.23}
\e
Of special interest is the case when
\f
q=\frac 12\left( n\pm 1\right) ~\left( n\mbox{ is odd}\right) \n{3.24}
\e
or
\f
q=\frac 12n~\left( n\mbox{ is even}\right) . \n{3.25}
\e
If all expressions (3.21) and/or (3.22) are zero for all values of $q\,$
with the exception of one or two ones defined by the condition (3.24) (or
(3.25)), the value $\xi ^{\widehat{i}}$ (or $\xi ^I$ ($\xi ^{I^{\prime }}))$
is called a fundamental h-spinor. Defining in a similar manner the
fundamental v-spinors we can introduce fundamental d-spinors as pairs of
fundamental h- and v-spinors. Here we remark that a h(v)-spinor $\xi ^{%
\widehat{i}}~(\xi ^{\widehat{a}})\,$ (we can also consider reduced
components) is always a fundamental one for $n(m)<7,$ which is a consequence
of (3.23)).

Finally, in this section, we note that the geometry of fundamental h- and
v-spinors is similar to that of usual fundamental spinors (see  Appendix
to the monograph [13]). We omit such details in this paper, but emphasize
that constructions with fundamental d-spinors, for a la-space, must be
adapted to the corresponding global splitting by N-connection of the space.
\vskip40pt
{\large \bf 4. D-spinor differential geometry}
\vskip15pt
The goal of the section is to formulate the differential geometry of
d-spinors on la-spaces.

We shall use denotations of type

$$
v^\alpha =(v^i,v^a)\in {{\cal \sigma}^\alpha } =({{\cal \sigma}
^i,{\cal \sigma}^a})\,\mbox{ and }\zeta ^{\underline{\alpha }}=(\zeta ^{%
\underline{i}},\zeta ^{\underline{a}})\in {{\cal \sigma}^{\underline{\alpha }}%
}=({{\cal \sigma} ^{\underline{i}}, {\cal \sigma}^{\underline{a}}})\,
$$
for, respectively, elements of modules of d-vector and irreduced d-spinor
fields (see details in [7]). D-tensors and d-spinor tensors (irreduced or
reduced) will be interpreted as elements of corresponding ${\cal \sigma }$
-modules, for instance,
$$
q_{~\beta ...}^\alpha \in {\cal \sigma ^\alpha ~_{\beta ....}}, \psi
_{~\underline{\beta }\quad ...}^{\underline{\alpha }\quad \underline{\gamma }%
}\in {\cal \sigma }_{~\underline{\beta }\quad ...}^{\underline{\alpha }\quad
\underline{\gamma }}~, \xi _{\quad JK^{\prime }N^{\prime
}}^{II^{\prime }}\in {\cal \sigma }_{\quad JK^{\prime }N^{\prime
}}^{II^{\prime }}~,...
$$

We can establish a correspondence between the la-adapted metric $g_{\alpha
\beta }$ (2.12) and d-spinor metric $\epsilon _{\underline{\alpha }%
\underline{\beta }}$ ( $\epsilon $-objects (3.8) for both h- and
v-subspaces of ${\cal E\,}$ ) of a la-space ${\cal E}$ by using the relation%
\f
g_{\alpha \beta }=-\frac 1{N(n)+N(m)}((\sigma _{(\alpha }(u))^{\underline{%
\alpha }_1\underline{\beta }_1}(\sigma _{\beta )}(u))^{\underline{\beta }_2%
\underline{\alpha }_2})\epsilon _{\underline{\alpha }_1\underline{\alpha }%
_2}\epsilon _{\underline{\beta }_1\underline{\beta }_2}, \n{4.1}
\e
where%
\f
(\sigma _\alpha (u))^{\underline{\nu }\underline{\gamma }}=l_\alpha ^{%
\widehat{\alpha }}(u)(\sigma _{\widehat{\alpha }})^{\underline{\nu }%
\underline{\gamma }}, \n{4.2}
\e
which is a consequence of formulas (3.1)-(3.4). In brief we can write (4.1)
as
\f
g_{\alpha \beta }=\epsilon _{\underline{\alpha }_1\underline{\alpha }%
_2}\epsilon _{\underline{\beta }_1\underline{\beta }_2} \n{4.3}
\e
if the $\sigma $-objects are considered as a fixed structure, whereas $%
\epsilon $-objects are treated as caring the metric ''dynamics '' , on
la-space. This variant is used, for instance, in the so-called 2-spinor
geometry [12,13] and should be preferred if we have to make explicit the
algebraic symmetry properties of d-spinor objects. An alternative way is to
considered as fixed the algebraic structure of $\epsilon $-objects and to
use variable components of $\sigma $-objects of type (4.2) for developing a
variational d-spinor approach to gravitational and matter field interactions
on la-spaces ( the spinor Ashtekar variables [24] are introduced in this
manner).

We note that a d-spinor metric
$$
\epsilon _{\underline{\nu }\underline{\tau }}=\left(
\begin{array}{cc}
\epsilon _{\underline{i}\underline{j}} & 0 \\
0 & \epsilon _{\underline{a}\underline{b}}
\end{array}
\right)
$$
on the d-spinor space ${\cal S}=({\cal S}_{(h)},{\cal S}_{(v)})$
can have symmetric or
antisymmetric h (v) -components $\epsilon _{\underline{i}\underline{j}}$ ($%
\epsilon _{\underline{a}\underline{b}})$ , see $\epsilon $-objects (3.8).
For simplicity, in this section (in order to avoid cumbersome calculations
connected with eight-fold periodicity on dimensions $n$ and $m$ of a la-space
${\cal E\ }$) we shall develop a general d-spinor formalism only by using
irreduced spinor spaces ${\cal S}_{(h)}$ and ${\cal S}_{(v)}.$
\vskip30pt
{\large \it 4.1. D-covariant derivation on la-spaces}
\vskip15pt
Let ${\cal E}$ be a la-space. We define the action on a d-spinor of a
d-covariant operator%
$$
\nabla _\alpha =\left( \nabla _i,\nabla _a\right) =(\sigma _\alpha )^{%
\underline{\alpha }_1\underline{\alpha }_2}\nabla _{^{\underline{\alpha }_1%
\underline{\alpha }_2}}=\left( (\sigma _i)^{\underline{i}_1\underline{i}%
_2}\nabla _{^{\underline{i}_1\underline{i}_2}},~(\sigma _a)^{\underline{a}_1%
\underline{a}_2}\nabla _{^{\underline{a}_1\underline{a}_2}}\right)
$$
(in brief, we shall write
$$
\nabla _\alpha =\nabla _{^{\underline{\alpha }_1\underline{\alpha }%
_2}}=\left( \nabla _{^{\underline{i}_1\underline{i}_2}},~\nabla _{^{%
\underline{a}_1\underline{a}_2}}\right)  )
$$
as a map
$$
\nabla _{{\underline{\alpha }}_1 {\underline{\alpha }}_2}\ :\ {\cal %
\sigma}^{\underline{\beta }}\rightarrow \sigma _\alpha ^{\underline{\beta }}=
\sigma _{{\underline{\alpha }}_1 {\underline{\alpha }}_2}^{\underline{\beta }}
$$
satisfying conditions%
$$
\nabla _\alpha (\xi ^{\underline{\beta }}+\eta ^{\underline{\beta }})=\nabla
_\alpha \xi ^{\underline{\beta }}+\nabla _\alpha \eta ^{\underline{\beta }},
$$
and%
$$
\nabla _\alpha (f\xi ^{\underline{\beta }})=f\nabla _\alpha \xi ^{\underline{%
\beta }}+\xi ^{\underline{\beta }}\nabla _\alpha f
$$
for every $\xi ^{\underline{\beta }},\eta ^{\underline{\beta }}\in {\cal %
\sigma ^{\underline{\beta }}}$ and $f$ being a scalar field on ${\cal E.\ }$
It is also required that one holds the Leibnitz rule%
$$
(\nabla _\alpha \zeta _{\underline{\beta }})\eta ^{\underline{\beta }%
}=\nabla _\alpha (\zeta _{\underline{\beta }}\eta ^{\underline{\beta }%
})-\zeta _{\underline{\beta }}\nabla _\alpha \eta ^{\underline{\beta }}
$$
and that $\nabla _\alpha \,$ is a real operator, i.e. it commuters with the
operation of complex conjugation:%
$$
\overline{\nabla _\alpha \psi _{\underline{\alpha }\underline{\beta }%
\underline{\gamma }...}}=\nabla _\alpha (\overline{\psi }_{\underline{\alpha
}\underline{\beta }\underline{\gamma }...}).
$$

Let now analyze the question on uniqueness of action on d-spinors of an
operator $\nabla _\alpha $ satisfying necessary conditions . Denoting by $%
\nabla _\alpha ^{(1)}$ and $\nabla _\alpha $ two such d-covariant operators
we consider the map%
\f
(\nabla _\alpha ^{(1)}-\nabla _\alpha ):{\cal \sigma ^{\underline{\beta }%
}\rightarrow \sigma _{\underline{\alpha }_1\underline{\alpha }_2}^{%
\underline{\beta }}}. \n{4.4}
\e
Because the action on a scalar $f$ of both operators $\nabla _\alpha ^{(1)}$
and $\nabla _\alpha $ must be identical, i.e.%
\f
\nabla _\alpha ^{(1)}f=\nabla _\alpha f, \n{4.5}
\e
the action (4.4) on $f=\omega _{\underline{\beta }}\xi ^{\underline{\beta }}$
must be written as
$$
(\nabla _\alpha ^{(1)}-\nabla _\alpha )(\omega _{\underline{\beta }}\xi ^{%
\underline{\beta }})=0.
$$
In consequence we conclude that there is an element $\Theta _{\underline{%
\alpha }_1\underline{\alpha }_2\underline{\beta }}^{\quad \quad \underline{%
\gamma }}\in {\cal \sigma }_{\underline{\alpha }_1\underline{\alpha }_2%
\underline{\beta }}^{\quad \quad \underline{\gamma }}$ for which%
\f
\nabla _{\underline{\alpha }_1\underline{\alpha }_2}^{(1)}\xi ^{\underline{%
\gamma }}=\nabla _{\underline{\alpha }_1\underline{\alpha }_2}\xi ^{%
\underline{\gamma }}+\Theta _{\underline{\alpha }_1\underline{\alpha }_2%
\underline{\beta }}^{\quad \quad \underline{\gamma }}\xi ^{\underline{\beta }%
} \n{4.6}
\e
and%
$$
\nabla _{\underline{\alpha }_1\underline{\alpha }_2}^{(1)}\omega _{%
\underline{\beta }}=\nabla _{\underline{\alpha }_1\underline{\alpha }%
_2}\omega _{\underline{\beta }}-\Theta _{\underline{\alpha }_1\underline{%
\alpha }_2\underline{\beta }}^{\quad \quad \underline{\gamma }}\omega _{%
\underline{\gamma }}~.
$$
The action of the operator (4.4) on a d-vector $v^\beta =v^{\underline{\beta
}_1\underline{\beta }_2}$ can be written by using formula (4.6) for both
indices $\underline{\beta }_1$ and $\underline{\beta }_2$ :%
$$
(\nabla _\alpha ^{(1)}-\nabla _\alpha )v^{\underline{\beta }_1\underline{%
\beta }_2}=\Theta _{\alpha \underline{\gamma }}^{\quad \underline{\beta }%
_1}v^{\underline{\gamma }\underline{\beta }_2}+\Theta _{\alpha \underline{%
\gamma }}^{\quad \underline{\beta }_2}v^{\underline{\beta }_1\underline{%
\gamma }}=
$$
$$
(\Theta _{\alpha \underline{\gamma }_1}^{\quad \underline{\beta }_1}\delta _{%
\underline{\gamma }_2}^{\quad \underline{\beta }_2}+\Theta _{\alpha
\underline{\gamma }_1}^{\quad \underline{\beta }_2}\delta _{\underline{%
\gamma }_2}^{\quad \underline{\beta }_1})v^{\underline{\gamma }_1\underline{%
\gamma }_2}=Q_{\ \alpha \gamma }^\beta v^\gamma ,
$$
where%
\f
Q_{\ \alpha \gamma }^\beta =Q_{\qquad \underline{\alpha }_1\underline{\alpha
}_2~\underline{\gamma }_1\underline{\gamma }_2}^{\underline{\beta }_1%
\underline{\beta }_2}=\Theta _{\alpha \underline{\gamma }_1}^{\quad
\underline{\beta }_1}\delta _{\underline{\gamma }_2}^{\quad \underline{\beta
}_2}+\Theta _{\alpha \underline{\gamma }_1}^{\quad \underline{\beta }%
_2}\delta _{\underline{\gamma }_2}^{\quad \underline{\beta }_1}. \n{4.7}
\e
The d-commutator $\nabla _{[\alpha }\nabla _{\beta ]}$ defines the d-torsion
(see (2.35) and (2.36)). So, applying operators $\nabla _{[\alpha
}^{(1)}\nabla _{\beta ]}^{(1)}$ and $\nabla _{[\alpha }\nabla _{\beta ]}$ on
$f=\omega _{\underline{\beta }}\xi ^{\underline{\beta }}$ we can write
$$
T_{\quad \alpha \beta }^{(1)\gamma }-T_{~\alpha \beta }^\gamma =Q_{~\beta
\alpha }^\gamma -Q_{~\alpha \beta }^\gamma
$$
with $Q_{~\alpha \beta }^\gamma $ from (4.7).

The action of operator $\nabla _\alpha ^{(1)}$ on d-spinor tensors of type $%
\chi _{\underline{\alpha }_1\underline{\alpha }_2\underline{\alpha }%
_3...}^{\qquad \quad \underline{\beta }_1\underline{\beta }_2...}$ must be
constructed by using formula (4.6) for every upper index $\underline{\beta }%
_1\underline{\beta }_2...$ and formula (4.7) for every lower index $%
\underline{\alpha }_1\underline{\alpha }_2\underline{\alpha }_3...$ .
\vskip30pt
{\large \it 4.2. Infeld - van der Waerden coefficients and d-connections}
\vskip15pt
Let $$\delta _{\underline{{\bf \alpha }}}^{\quad \underline{\alpha }}=\left(
\delta _{\underline{{\bf 1}}}^{\quad \underline{i}},\delta _{\underline{{\bf %
2}}}^{\quad \underline{i}},...,\delta _{\underline{{\bf N(n)}}}^{\quad
\underline{i}},\delta _{\underline{{\bf 1}}}^{\quad \underline{a}},\delta _{%
\underline{{\bf 2}}}^{\quad \underline{a}},...,\delta _{\underline{{\bf N(m)}%
}}^{\quad \underline{i}}\right) $$
be a d-spinor basis. The dual to it basis
is denoted as
$$\delta _{\underline{\alpha }}^{\quad \underline{{\bf \alpha }}%
}=\left( \delta _{\underline{i}}^{\quad \underline{{\bf 1}}},\delta _{%
\underline{i}}^{\quad \underline{{\bf 2}}},...,\delta _{\underline{i}%
}^{\quad \underline{{\bf N(n)}}},\delta _{\underline{i}}^{\quad \underline{%
{\bf 1}}},\delta _{\underline{i}}^{\quad \underline{{\bf 2}}},...,\delta _{%
\underline{i}}^{\quad \underline{{\bf N(m)}}}\right) .$$
 A d-spinor $\kappa ^{%
\underline{\alpha }}\in {\cal \sigma }$ $^{\underline{\alpha }}$ has
components $\kappa ^{\underline{{\bf \alpha }}}=\kappa ^{\underline{\alpha }%
}\delta _{\underline{\alpha }}^{\quad \underline{{\bf \alpha }}}.$ Taking
into account that
$$
\delta _{\underline{{\bf \alpha }}}^{\quad \underline{\alpha }}\delta _{%
\underline{{\bf \beta }}}^{\quad \underline{\beta }}\nabla _{\underline{%
\alpha }\underline{\beta }}=\nabla _{\underline{{\bf \alpha }}\underline{%
{\bf \beta }}},
$$
we write out the components $\nabla _{\underline{\alpha }\underline{\beta }}$
$\kappa ^{\underline{\gamma }}$ as%
$$
\delta _{\underline{{\bf \alpha }}}^{\quad \underline{\alpha }}~\delta _{%
\underline{{\bf \beta }}}^{\quad \underline{\beta }}~\delta _{\underline{%
\gamma }}^{\quad \underline{{\bf \gamma }}}~\nabla _{\underline{\alpha }%
\underline{\beta }}\kappa ^{\underline{\gamma }}=\delta _{\underline{{\bf %
\epsilon }}}^{\quad \underline{\tau }}~\delta _{\underline{\tau }%
}^{\quad \underline{{\bf \gamma }}}~\nabla _{\underline{{\bf \alpha }}%
\underline{{\bf \beta }}}\kappa ^{\underline{{\bf \epsilon }}}+\kappa ^{%
\underline{{\bf \epsilon }}}~\delta _{\underline{\epsilon }}^{\quad
\underline{{\bf \gamma }}}~\nabla _{\underline{{\bf \alpha }}\underline{{\bf %
\beta }}}\delta _{\underline{{\bf \epsilon }}}^{\quad \underline{\epsilon }%
}=
$$
\f
\nabla _{\underline{{\bf \alpha }}\underline{{\bf \beta }}}\kappa ^{%
\underline{{\bf \gamma }}}+\kappa ^{\underline{{\bf \epsilon }}}\gamma _{~%
\underline{{\bf \alpha }}\underline{{\bf \beta }}\underline{{\bf \epsilon }}%
}^{\underline{{\bf \gamma }}},\n{4.8}
\e
where the coordinate components of the d-spinor connection $\gamma _{~%
\underline{{\bf \alpha }}\underline{{\bf \beta }}\underline{{\bf \epsilon }}%
}^{\underline{{\bf \gamma }}}$ are defined as
\f
\gamma _{~\underline{{\bf \alpha }}\underline{{\bf \beta }}\underline{{\bf %
\epsilon }}}^{\underline{{\bf \gamma }}}\doteq \delta _{\underline{\tau }%
}^{\quad \underline{{\bf \gamma }}}~\nabla _{\underline{{\bf \alpha }}%
\underline{{\bf \beta }}}\delta _{\underline{{\bf \epsilon }}}^{\quad
\underline{\tau }}. \n{4.9}
\e

We call the Infeld - van der Waerden d-symbols a set of $\sigma $-objects ($%
\sigma _{{\bf \alpha }})^{\underline{{\bf \alpha }}\underline{{\bf \beta }}}$
parametrized with respect to a coordinate d-spinor basis. Defining
$$
\nabla _{{\bf \alpha }}=(\sigma _{{\bf \alpha }})^{\underline{{\bf \alpha }}%
\underline{{\bf \beta }}}~\nabla _{\underline{{\bf \alpha }}\underline{{\bf %
\beta }}},
$$
introducing denotations
$$
\gamma ^{\underline{{\bf \gamma }}}{}_{{\bf \alpha \underline{\tau }}}\doteq
\gamma ^{\underline{{\bf \gamma }}}{}_{{\bf \underline{\alpha }\underline{%
\beta }\underline{\tau }}}~(\sigma _{{\bf \alpha }})^{\underline{{\bf \alpha
}}\underline{{\bf \beta }}}
$$
and using properties (4.8) we can write relations%
\f
l_{{\bf \alpha }}^\alpha ~\delta _{\underline{\beta }}^{\quad \underline{%
{\bf \beta }}}~\nabla _\alpha \kappa ^{\underline{\beta }}=\nabla _{{\bf %
\alpha }}\kappa ^{\underline{{\bf \beta }}}+\kappa ^{\underline{{\bf \delta }%
}}~\gamma _{~{\bf \alpha }\underline{{\bf \delta }}}^{\underline{{\bf \beta }%
}} \n{4.10}
\e
and%
\f
l_{{\bf \alpha }}^\alpha ~\delta _{\underline{{\bf \beta }}}^{\quad
\underline{\beta }}~\nabla _\alpha ~\mu _{\underline{\beta }}=\nabla _{{\bf %
\alpha }}~\mu _{\underline{{\bf \beta }}}-\mu _{\underline{{\bf \delta }}%
}\gamma _{~{\bf \alpha }\underline{{\bf \beta }}}^{\underline{{\bf \delta }}%
} \n{4.11}
\e
for d-covariant derivations $~\nabla _{\underline{\alpha }}\kappa ^{%
\underline{\beta }}$ and $\nabla _{\underline{\alpha }}~\mu _{\underline{%
\beta }}.$

We can consider expressions similar to (4.10) and (4.11) for values having
both types of d-spinor and d-tensor indices, for instance,%
$$
l_{{\bf \alpha }}^\alpha ~l_\gamma ^{{\bf \gamma }}~\delta _{\underline{{\bf %
\delta }}}^{\quad \underline{\delta }}~\nabla _\alpha \theta _{\underline{%
\delta }}^{~\gamma }=\nabla _{{\bf \alpha }}\theta _{\underline{{\bf \delta }%
}}^{~{\bf \gamma }}-\theta _{\underline{{\bf \epsilon }}}^{~{\bf \gamma }%
}\gamma _{~\underline{{\bf \alpha }}\underline{{\bf \delta }}}^{\underline{%
{\bf \epsilon }}}+\theta _{\underline{{\bf \delta }}}^{~{\bf \tau }}~\Gamma
_{\quad {\bf \alpha \tau }}^{~{\bf \gamma }}
$$
(we can prove this by a straightforward calculation of the derivation $%
\nabla _\alpha (\theta _{\underline{{\bf \epsilon }}}^{~{\bf \tau }}$ $%
~\delta _{\underline{\delta }}^{\quad \underline{{\bf \epsilon }}}~l_{{\bf %
\tau }}^\gamma )$ ).

Now we shall consider some possible relations between components of
d-connections $\gamma _{~\underline{{\bf \alpha }}\underline{{\bf \delta }}%
}^{\underline{{\bf \epsilon }}}$ and $\Gamma _{\quad {\bf \alpha \tau }}^{~%
{\bf \gamma }}$ and derivations of $(\sigma _{{\bf \alpha }})^{\underline{%
{\bf \alpha }}\underline{{\bf \beta }}}$ . According to definitions (2.12)
we can write
$$
\Gamma _{~{\bf \beta \gamma }}^{{\bf \alpha }}=l_\alpha ^{{\bf \alpha }%
}\nabla _{{\bf \gamma }}l_{{\bf \beta }}^\alpha =l_\alpha ^{{\bf \alpha }%
}\nabla _{{\bf \gamma }}(\sigma _{{\bf \beta }})^{\underline{\epsilon }
\underline{\tau }}=l_\alpha ^{{\bf \alpha }}\nabla _{{\bf \gamma }}((\sigma
_{{\bf \beta }})^{\underline{{\bf \epsilon }}\underline{{\bf \tau }}}\delta
_{\underline{{\bf \epsilon }}}^{~\underline{\epsilon }}\delta _{\underline{%
{\bf \tau }}}^{~\underline{\tau }})=
$$
$$
l_\alpha ^{{\bf \alpha }}\delta _{\underline{{\bf \alpha }}}^{~\underline{%
\alpha }}\delta _{\underline{{\bf \epsilon }}}^{~\underline{\epsilon }%
}\nabla _{{\bf \gamma }}(\sigma _{{\bf \beta }})^{\underline{{\bf \alpha }}
\underline{{\bf \epsilon }}}+l_\alpha ^{{\bf \alpha }}(\sigma _{{\bf \beta }%
})^{\underline{{\bf \epsilon }}\underline{{\bf \tau }}}(\delta _{\underline{%
{\bf \tau }}}^{~\underline{\tau }}\nabla _{{\bf \gamma }}\delta _{\underline{%
{\bf \epsilon }}}^{~\underline{\epsilon }}+\delta _{\underline{{\bf \epsilon
}}}^{~\underline{\epsilon }}\nabla _{{\bf \gamma }}\delta _{\underline{{\bf %
\tau }}}^{~\underline{\tau }})=
$$
$$
l_{\underline{{\bf \epsilon }}\underline{{\bf \tau }}}^{{\bf \alpha }%
}~\nabla _{{\bf \gamma }}(\sigma _{{\bf \beta }})^{\underline{{\bf \epsilon }%
}\underline{{\bf \tau }}}+l_{\underline{{\bf \mu }}\underline{{\bf \nu }}}^{%
{\bf \alpha }}\delta _{\underline{\epsilon }}^{~\underline{{\bf \mu }}%
}\delta _{\underline{\tau }}^{~\underline{{\bf \nu }}}(\sigma _{{\bf \beta }%
})^{\underline{\epsilon }\underline{\tau }}(\delta _{\underline{{\bf \tau }}%
}^{~\underline{\tau }}\nabla _{{\bf \gamma }}\delta _{\underline{{\bf %
\epsilon }}}^{~\underline{\epsilon }}+\delta _{\underline{{\bf \epsilon }}%
}^{~\underline{\epsilon }}\nabla _{{\bf \gamma }}\delta _{\underline{{\bf %
\tau }}}^{~\underline{\tau }}),
$$
where $l_{\alpha}^{{\bf \alpha }%
}=(\sigma _{\underline{{\bf \epsilon }}\underline{{\bf \tau }}})^{{\bf %
\alpha }}$ , from which it follows%
$$
(\sigma _{{\bf \alpha }})^{\underline{{\bf \mu }}\underline{{\bf \nu }}%
}(\sigma _{\underline{{\bf \alpha }}\underline{{\bf \beta }}})^{{\bf \beta }%
}\Gamma _{~{\bf \gamma \beta }}^{{\bf \alpha }}= (\sigma _{\underline{{\bf %
\alpha }}\underline{{\bf \beta }}})^{{\bf \beta }}\nabla _{{\bf \gamma }%
}(\sigma _{{\bf \alpha }})^{\underline{{\bf \mu }}\underline{{\bf \nu }}%
}+\delta _{\underline{{\bf \beta }}}^{~\underline{{\bf \nu }}}\gamma _{~{\bf %
\gamma \underline{\alpha }}}^{\underline{{\bf \mu }}}+\delta _{\underline{%
{\bf \alpha }}}^{~\underline{{\bf \mu }}}\gamma _{~{\bf \gamma \underline{%
\beta }}}^{\underline{{\bf \nu }}}.
$$
Connecting the last expression on \underline{${\bf \beta }$} and \underline{$%
{\bf \nu }$} and using an orthonormalized d-spinor basis when $\gamma _{~%
{\bf \gamma \underline{\beta }}}^{\underline{{\bf \beta }}}=0$ (a
consequence from (4.9)) we have
\f
\gamma _{~{\bf \gamma \underline{\alpha }}}^{\underline{{\bf \mu }}}=\frac
1{N(n)+N(m)}(\Gamma _{\quad {\bf \gamma ~\underline{\alpha }\underline{\beta
}}}^{\underline{{\bf \mu }}\underline{{\bf \beta }}}-(\sigma _{\underline{%
{\bf \alpha }}\underline{{\bf \beta }}})^{{\bf \beta }}\nabla _{{\bf \gamma }%
}(\sigma _{{\bf \beta }})^{\underline{{\bf \mu }}\underline{{\bf \beta }}%
}), \n{4.12}
\e
where
\f
\Gamma _{\quad {\bf \gamma ~\underline{\alpha }\underline{\beta }}}^{%
\underline{{\bf \mu }}\underline{{\bf \beta }}}=(\sigma _{{\bf \alpha }})^{%
\underline{{\bf \mu }}\underline{{\bf \beta }}}(\sigma _{\underline{{\bf %
\alpha }}\underline{{\bf \beta }}})^{{\bf \beta }}\Gamma _{~{\bf \gamma
\beta }}^{{\bf \alpha }}. \n{4.13}
\e
We also note here that, for instance, for the canonical and Berwald
connections, Christoffel d-symbols we can express d-spinor connection (4.13)
through corresponding locally adapted derivations of components of metric
and N-connection by introducing respectively coefficients (2.22) and (2.20),
or (2.23) instead of $\Gamma _{~{\bf \gamma \beta }}^{{\bf \alpha }}$ in
(4.13) and than in (4.12).
\vskip35pt
{\large \it 4.3. D-spinors of la-space curvature and torsion}
\vskip15pt
The d-tensor indices of the commutator (2.35), $\Delta _{\alpha \beta },$
can be transformed into d-spinor ones:%
\f
\Box _{\underline{\alpha }\underline{\beta }}=(\sigma ^{\alpha \beta })_{%
\underline{\alpha }\underline{\beta }}\Delta _{\alpha \beta }=(\Box _{%
\underline{i}\underline{j}},\Box _{\underline{a}\underline{b}}), \n{4.14}
\e
with h- and v-components%
$$
\Box _{\underline{i}\underline{j}}=(\sigma ^{\alpha \beta })_{\underline{i}%
\underline{j}}\Delta _{\alpha \beta }\text{ and }\Box _{\underline{a}%
\underline{b}}=(\sigma ^{\alpha \beta })_{\underline{a}\underline{b}}\Delta
_{\alpha \beta }
$$
being symmetric or antisymmetric in dependence of corresponding values of
dimensions $n\,$ and $m$ (see eight-fold parametizations (3.18),\ (3.19) and
(3.20)). Considering the actions of operator (4.14) on d-spinors $\pi ^{%
\underline{\gamma }}$ and $\mu _{\underline{\gamma }}$ we introduce
the d-spinor curvature $X_{\underline{\delta }\quad \underline{\alpha }%
\underline{\beta }}^{\quad \underline{\gamma }}\,$ as to satisfy equations%
\f
\Box _{\underline{\alpha }\underline{\beta }}\ \pi ^{\underline{\gamma }}=X_{%
\underline{\delta }\quad \underline{\alpha }\underline{\beta }}^{\quad
\underline{\gamma }}\pi ^{\underline{\delta }} \n{4.15}
\e
and%
$$
\Box _{\underline{\alpha }\underline{\beta }}\ \mu _{\underline{\gamma }%
}=X_{\underline{\gamma }\quad \underline{\alpha }\underline{\beta }%
}^{\quad \underline{\delta }}\mu _{\underline{\delta }}.
$$
The gravitational d-spinor $\Psi _{\underline{\alpha }\underline{\beta }%
\underline{\gamma }\underline{\delta }}$ is defined by a corresponding
symmetrization of d-spinor indices:%
\f
\Psi _{\underline{\alpha }\underline{\beta }\underline{\gamma }\underline{%
\delta }}=X_{(\underline{\alpha }|\underline{\beta }|\underline{\gamma }%
\underline{\delta })}. \n{4.16}
\e
We note that d-spinor tensors $X_{\underline{\delta }\quad \underline{\alpha
}\underline{\beta }}^{\quad \underline{\gamma }}$ and $\Psi _{\underline{%
\alpha }\underline{\beta }\underline{\gamma }\underline{\delta }}\,$ are
transformed into similar 2-spinor objects on locally isotropic spaces
[12,13] if we consider vanishing of the N-connection structure and a limit
to a locally isotropic space.

Putting $\delta _{\underline{\gamma }}^{\quad {\bf \underline{\gamma }}}$
instead of $\mu _{\underline{\gamma }}$ in (4.15) and using (4.16)
we can express respectively the curvature and gravitational d-spinors as
$$
X_{\underline{\gamma }\underline{\delta }\underline{\alpha }\underline{\beta
}}=\delta _{\underline{\delta }\underline{{\bf \tau }}}\Box _{\underline{%
\alpha }\underline{\beta }}\delta _{\underline{\gamma }}^{\quad {\bf
\underline{\tau }}}
$$
and%
$$
\Psi _{\underline{\gamma }\underline{\delta }\underline{\alpha }\underline{%
\beta }}=\delta _{\underline{\delta }\underline{{\bf \tau }}}\Box _{(%
\underline{\alpha }\underline{\beta }}\delta _{\underline{\gamma })}^{\quad
{\bf \underline{\tau }}}.
$$

The d-spinor torsion $T_{\qquad \underline{\alpha }\underline{\beta }}^{%
\underline{\gamma }_1\underline{\gamma }_2}$ is defined similarly as for
d-tensors (see (2.36)) by using the d-spinor commutator (4.15) and equations
\f
\Box _{\underline{\alpha }\underline{\beta }}f=T_{\qquad \underline{\alpha }%
\underline{\beta }}^{\underline{\gamma }_1\underline{\gamma }_2}\nabla _{%
\underline{\gamma }_1\underline{\gamma }_2}f. \n{4.17}
\e

The d-spinor components $R_{\underline{\gamma }_1\underline{\gamma }_2\qquad
\underline{\alpha }\underline{\beta }}^{\qquad \underline{\delta }_1%
\underline{\delta }_2}$ of the curvature d-tensor $R_{\gamma \quad \alpha
\beta }^{\quad \delta }$ can be computed by using relations (4.14), and
(4.15) and (4.17) as to satisfy the equations (the d-spinor analogous of
equations (2.37) )%
\f
(\Box _{\underline{\alpha }\underline{\beta }}-T_{\qquad \underline{\alpha }%
\underline{\beta }}^{\underline{\gamma }_1\underline{\gamma }_2}\nabla _{%
\underline{\gamma }_1\underline{\gamma }_2})V^{\underline{\delta }_1%
\underline{\delta }_2}=R_{\underline{\gamma }_1\underline{\gamma }_2\qquad
\underline{\alpha }\underline{\beta }}^{\qquad \underline{\delta }_1%
\underline{\delta }_2}V^{\underline{\gamma }_1\underline{\gamma }_2}, \n{4.18}
\e
where d-vector $V^{\underline{\gamma }_1\underline{\gamma }_2}$ is
considered as a product of d-spinors, i.e. $V^{\underline{\gamma }_1%
\underline{\gamma }_2}=\nu ^{\underline{\gamma }_1}\mu ^{\underline{\gamma }%
_2}$. We find

$$
R_{\underline{\gamma }_1\underline{\gamma }_2\qquad \underline{\alpha }%
\underline{\beta }}^{\qquad \underline{\delta }_1\underline{\delta }%
_2}=\left( X_{\underline{\gamma }_1~\underline{\alpha }\underline{\beta }%
}^{\quad \underline{\delta }_1}+T_{\qquad \underline{\alpha }\underline{%
\beta }}^{\underline{\tau }_1\underline{\tau }_2}\quad \gamma _{\quad
\underline{\tau }_1\underline{\tau }_2\underline{\gamma }_1}^{\underline{%
\delta }_1}\right) \delta _{\underline{\gamma }_2}^{\quad \underline{\delta }%
_2}+$$    \f
\left( X_{\underline{\gamma }_2~\underline{\alpha }\underline{\beta }%
}^{\quad \underline{\delta }_2}+T_{\qquad \underline{\alpha }\underline{%
\beta }}^{\underline{\tau }_1\underline{\tau }_2}\quad \gamma _{\quad
\underline{\tau }_1\underline{\tau }_2\underline{\gamma }_2}^{\underline{%
\delta }_2}\right) \delta _{\underline{\gamma }_1}^{\quad \underline{\delta }%
_1}. \n{4.19}
\e

It is convenient to use this d-spinor expression for the curvature d-tensor
$$
R_{\underline{\gamma }_1\underline{\gamma }_2\qquad \underline{\alpha }_1%
\underline{\alpha }_2\underline{\beta }_1\underline{\beta }_2}^{\qquad
\underline{\delta }_1\underline{\delta }_2}=\left( X_{\underline{\gamma }_1~%
\underline{\alpha }_1\underline{\alpha }_2\underline{\beta }_1\underline{%
\beta }_2}^{\quad \underline{\delta }_1}+T_{\qquad \underline{\alpha }_1%
\underline{\alpha }_2\underline{\beta }_1\underline{\beta }_2}^{\underline{%
\tau }_1\underline{\tau }_2}~\gamma _{\quad \underline{\tau }_1\underline{%
\tau }_2\underline{\gamma }_1}^{\underline{\delta }_1}\right) \delta _{%
\underline{\gamma }_2}^{\quad \underline{\delta }_2}+$$
$$\left( X_{\underline{%
\gamma }_2~\underline{\alpha }_1\underline{\alpha }_2\underline{\beta }_1%
\underline{\beta }_2}^{\quad \underline{\delta }_2}+T_{\qquad \underline{%
\alpha }_1\underline{\alpha }_2\underline{\beta }_1\underline{\beta }_2~}^{%
\underline{\tau }_1\underline{\tau }_2}\gamma _{\quad \underline{\tau }_1%
\underline{\tau }_2\underline{\gamma }_2}^{\underline{\delta }_2}\right)
\delta _{\underline{\gamma }_1}^{\quad \underline{\delta }_1}
$$
in order to get the d-spinor components of the Ricci d-tensor%
$$
R_{\underline{\gamma }_1\underline{\gamma }_2\underline{\alpha }_1\underline{%
\alpha }_2}=R_{\underline{\gamma }_1\underline{\gamma }_2\qquad \underline{%
\alpha }_1\underline{\alpha }_2\underline{\delta }_1\underline{\delta }%
_2}^{\qquad \underline{\delta }_1\underline{\delta }_2}=
$$
\f
X_{\underline{\gamma }_1~\underline{\alpha }_1\underline{\alpha }_2%
\underline{\delta }_1\underline{\gamma }_2}^{\quad \underline{\delta }%
_1}+T_{\qquad \underline{\alpha }_1\underline{\alpha }_2\underline{\delta }_1%
\underline{\gamma }_2}^{\underline{\tau }_1\underline{\tau }_2}~\gamma
_{\quad \underline{\tau }_1\underline{\tau }_2\underline{\gamma }_1}^{%
\underline{\delta }_1}+X_{\underline{\gamma }_2~\underline{\alpha }_1%
\underline{\alpha }_2\underline{\delta }_1\underline{\gamma }_2}^{\quad
\underline{\delta }_2}+T_{\qquad \underline{\alpha }_1\underline{\alpha }_2%
\underline{\gamma }_1\underline{\delta }_2~}^{\underline{\tau }_1\underline{%
\tau }_2}\gamma _{\quad \underline{\tau }_1\underline{\tau }_2\underline{%
\gamma }_2}^{\underline{\delta }_2} \n{4.20}
\e
and this d-spinor decomposition of the scalar curvature:%
$$
\overleftarrow{R}=R_{\qquad \underline{\alpha }_1\underline{\alpha }_2}^{%
\underline{\alpha }_1\underline{\alpha }_2}=X_{\quad ~\underline{~\alpha }%
_1\quad \underline{\delta }_1\underline{\alpha }_2}^{\underline{\alpha }_1%
\underline{\delta }_1~~\underline{\alpha }_2}+T_{\qquad ~~\underline{\alpha }%
_2\underline{\delta }_1}^{\underline{\tau }_1\underline{\tau }_2\underline{%
\alpha }_1\quad ~\underline{\alpha }_2}~\gamma _{\quad \underline{\tau }_1%
\underline{\tau }_2\underline{\alpha }_1}^{\underline{\delta }_1}+$$
\f X_{\qquad
\quad \underline{\alpha }_2\underline{\delta }_2\underline{\alpha }_1}^{%
\underline{\alpha }_2\underline{\delta }_2\underline{\alpha }_1}+T_{\qquad
\underline{\alpha }_1\quad ~\underline{\delta }_2~}^{\underline{\tau }_1%
\underline{\tau }_2~~\underline{\alpha }_2\underline{\alpha }_1}\gamma
_{\quad \underline{\tau }_1\underline{\tau }_2\underline{\alpha }_2}^{%
\underline{\delta }_2}. \n{4.21}
\e

Putting (4.20) and (4.21) into (2.42) and, correspondingly, (2.43) we find
the d-spinor components of the Einstein and $\Phi _{\alpha \beta }$
d-tensors:%
$$
\overleftarrow{G}_{\gamma \alpha }=\overleftarrow{G}_{\underline{\gamma }_1%
\underline{\gamma }_2\underline{\alpha }_1\underline{\alpha }_2}=X_{%
\underline{\gamma }_1~\underline{\alpha }_1\underline{\alpha }_2\underline{%
\delta }_1\underline{\gamma }_2}^{\quad \underline{\delta }_1}+T_{\qquad
\underline{\alpha }_1\underline{\alpha }_2\underline{\delta }_1\underline{%
\gamma }_2}^{\underline{\tau }_1\underline{\tau }_2}~\gamma _{\quad
\underline{\tau }_1\underline{\tau }_2\underline{\gamma }_1}^{\underline{%
\delta }_1}+$$
$$X_{\underline{\gamma }_2~\underline{\alpha }_1\underline{\alpha }%
_2\underline{\delta }_1\underline{\gamma }_2}^{\quad \underline{\delta }%
_2}+T_{\qquad \underline{\alpha }_1\underline{\alpha }_2\underline{\gamma }_1%
\underline{\delta }_2~}^{\underline{\tau }_1\underline{\tau }_2}\gamma
_{\quad \underline{\tau }_1\underline{\tau }_2\underline{\gamma }_2}^{%
\underline{\delta }_2}-
$$
$$
\frac 12\varepsilon _{\underline{\gamma }_1\underline{\alpha }_1}\varepsilon
_{\underline{\gamma }_2\underline{\alpha }_2} [ X_{\quad ~\underline{%
~\beta }_1\quad \underline{\mu }_1\underline{\beta }_2}^{\underline{\beta }_1%
\underline{\mu }_1~~\underline{\beta }_2}+T_{\qquad ~~\underline{\beta }_2%
\underline{\mu }_1}^{\underline{\tau }_1\underline{\tau }_2\underline{\beta }%
_1\quad ~\underline{\beta }_2}~\gamma _{\quad \underline{\tau }_1\underline{%
\tau }_2\underline{\beta }_1}^{\underline{\mu }_1}+$$
\f X_{\qquad \quad
\underline{\beta }_2\underline{\mu }_2\underline{\nu }_1}^{\underline{\beta }%
_2\underline{\mu }_2\underline{\nu }_1}+T_{\qquad \underline{\beta }_1\quad ~%
\underline{\delta }_2~}^{\underline{\tau }_1\underline{\tau }_2~~\underline{%
\beta }_2\underline{\beta }_1}\gamma _{\quad \underline{\tau }_1\underline{%
\tau }_2\underline{\beta }_2}^{\underline{\delta }_2} ] \n{4.22}
\e
and%
$$
\Phi _{\gamma \alpha }=\Phi _{\underline{\gamma }_1\underline{\gamma }_2%
\underline{\alpha }_1\underline{\alpha }_2}=
\frac 1{2(n+m)}\varepsilon _{\underline{\gamma }_1\underline{\alpha }%
_1}\varepsilon _{\underline{\gamma }_2\underline{\alpha }_2} [ X_{\quad ~%
\underline{~\beta }_1\quad \underline{\mu }_1\underline{\beta }_2}^{%
\underline{\beta }_1\underline{\mu }_1~~\underline{\beta }_2}+T_{\qquad ~~%
\underline{\beta }_2\underline{\mu }_1}^{\underline{\tau }_1\underline{\tau }%
_2\underline{\beta }_1\quad ~\underline{\beta }_2}~\gamma _{\quad \underline{%
\tau }_1\underline{\tau }_2\underline{\beta }_1}^{\underline{\mu }%
_1}+$$
$$X_{\qquad \quad \underline{\beta }_2\underline{\mu }_2\underline{\nu }%
_1}^{\underline{\beta }_2\underline{\mu }_2\underline{\nu }_1}+T_{\qquad
\underline{\beta }_1\quad ~\underline{\delta }_2~}^{\underline{\tau }_1%
\underline{\tau }_2~~\underline{\beta }_2\underline{\beta }_1}\gamma _{\quad
\underline{\tau }_1\underline{\tau }_2\underline{\beta }_2}^{\underline{%
\delta }_2} ] -
$$
$$
\frac 12 [ X_{\underline{\gamma }_1~\underline{\alpha }_1\underline{%
\alpha }_2\underline{\delta }_1\underline{\gamma }_2}^{\quad \underline{%
\delta }_1}+T_{\qquad \underline{\alpha }_1\underline{\alpha }_2\underline{%
\delta }_1\underline{\gamma }_2}^{\underline{\tau }_1\underline{\tau }%
_2}~\gamma _{\quad \underline{\tau }_1\underline{\tau }_2\underline{\gamma }%
_1}^{\underline{\delta }_1} + $$
\f
X_{\underline{\gamma }_2~\underline{\alpha }_1%
\underline{\alpha }_2\underline{\delta }_1\underline{\gamma }_2}^{\quad
\underline{\delta }_2}+T_{\qquad \underline{\alpha }_1\underline{\alpha }_2%
\underline{\gamma }_1\underline{\delta }_2~}^{\underline{\tau }_1\underline{%
\tau }_2}\gamma _{\quad \underline{\tau }_1\underline{\tau }_2\underline{%
\gamma }_2}^{\underline{\delta }_2} ] . \n{4.23}
\e

The components of the conformal Weyl d-spinor can be computed by putting
d-spinor values of the curvature (4.19) and Ricci (4.20) d-tensors into
corresponding expression for the d-tensor (2.40). We omit this calculus in
this work.
\vskip40pt
 {\large \bf 5. Field d-tensor and d-spinor equations on la-spaces}
\vskip15pt
The problem of formulation gravitational and gauge field equations on
different types of la-spaces is considered, for instance, in [2,4,5] and [9].
 In this section we shall introduce the basic field equations for
gravitational and matter field la-interactions in a generalized form for
generic la-spaces.
\vskip30pt
{\large \it 5.1. Locally anisotropic scalar field interactions}
\vskip15pt
Let $\varphi \left( u\right) =(\varphi _1\left( u\right) ,\varphi _2\left(
u\right) \dot ,...,\varphi _k\left( u\right) )$ be a complex k-component
scalar field of mass $\mu $ on la-space ${\cal E.}$ The d-covariant
generalization of the conformally invariant (in the massless case) scalar
field equation [11,12] can be defined by using the d'Alambert locally
anisotropic operator [25,26] $\Box =D^\alpha D_\alpha $, where $D_\alpha $
is a d-covariant derivation on ${\cal E}$ satisfying conditions (2.14) and
(2.15):%

\f
(\Box +\frac{n+m-2}{4(n+m-1)}\overleftarrow{R}+\mu ^2)\varphi \left(
u\right) =0. \n{5.1}
\e
We must change d-covariant derivation $D_{\alpha }$ into $%
^{\diamond }D_\alpha =D_\alpha +ieA_\alpha $ and take into account the
d-vector current
$$
J_\alpha ^{(0)}\left( u\right) =i(\left( \overline \varphi  \left( u\right)
D_\alpha \varphi \left( u\right) -D_\alpha \overline \varphi \left( u\right)
)\varphi \left( u\right) \right)
$$
if interactions between locally anisotropic electromagnetic field ( d-vector
potential $A_\alpha $ ), where $e$ is the electromagnetic constant, and
charged scalar field $\varphi $ are considered. The equations (5.1) are
(locally adapted to the N-connection structure) Euler equations for the
Lagrangian%
\f
{\cal L}^{(0)}\left( u\right) =\sqrt{|g|}\left[ g^{\alpha \beta }\delta
_\alpha \overline \varphi \left( u\right) \delta _\beta \varphi \left( u\right)
-\left( \mu ^2+\frac{n+m-2}{4(n+m-1)}\right) \overline \varphi \left( u\right)
\varphi \left( u\right) \right], \n{5.2}
\e
where $|g|=det g_{\alpha \beta}.$

The locally adapted variations of the action with Lagrangian (5.2) on
variables $\varphi \left( u\right) $ and $\overline \varphi \left( u\right) $
leads to the locally anisotropic generalization of the energy-momentum
tensor,%
\f
E_{\alpha \beta }^{(0,can)}\left( u\right) =\delta _\alpha \overline \varphi
 \left( u\right) \delta _\beta \varphi \left( u\right) +\delta _\beta
\overline \varphi \left( u\right) \delta _\alpha \varphi \left( u\right) -
\frac 1{%
\sqrt{|g|}}g_{\alpha \beta }{\cal L}^{(0)}\left( u\right) , \n{5.3}
\e
and a similar variation on the components of a d-metric (2.12) leads to a
symmetric metric energy-momentum d-tensor,%
\f
E_{\alpha \beta }^{(0)}\left( u\right) =E_{(\alpha \beta )}^{(0,can)}\left(
u\right) -\frac{n+m-2}{2(n+m-1)}\left[ R_{(\alpha \beta )}+D_{(\alpha
}D_{\beta )}-g_{\alpha \beta }\Box \right] \overline \varphi \left( u\right)
\varphi \left( u\right) . \n{5.4}
\e
Here we note that we can obtain a nonsymmetric energy-momentum d-tensor if
we firstly vary on $G_{\alpha \beta }$ and than impose constraints of type
(2.10) in order to have a compatibility with the N-connection structure. We
also conclude that the existence of a N-connection in v-bundle ${\cal E}$
results in a nonequivalence of energy-momentum d-tensors (5.3) and (5.4),
nonsymmetry of the Ricci tensor (see (2.33)), nonvanishing of the
d-covariant derivation of the Einstein d-tensor (4.23), $D_\alpha
\overleftarrow{G}^{\alpha \beta }\neq 0$ and, in consequence, a
corresponding breaking of conservation laws on la-spaces when $D_\alpha
E^{\alpha \beta }\neq 0\,$  [2]. The problem of formulation
of conservation laws on la-spaces is discussed in details and two variants
of its solution (by using nearly autoparallel maps and tensor integral
formalism on la-multispaces) are proposed in [26]. In this work we shall
present only straightforward generalizations of field equations and
necessary formulas for energy-momentum d-tensors of matter fields on ${\cal E%
}$ considering that it is naturally that the conservation laws (usually
being consequences of global, local and/or intrinsic symmetries of the
fundamental space-time and of the type of field interactions) have to be
broken on locally anisotropic spaces.
\vskip30pt
{\large \em 5.2 Proca equations on la-spaces}
\vskip15pt
Let consider a d-vector Proca field $\varphi _\alpha \left( u\right) $
with mass $\mu ^2$ (locally anisotropic Proca field ) interacting with
exterior la-gravitational field. From the Lagrangian
\f
{\cal L}^{(1)}\left( u\right) =\sqrt{\left| g\right| }\left[ -\frac
12 {\overline f}_{\alpha \beta }\left( u\right)
 f^{\alpha \beta }\left( u\right) +\mu
^2 {\overline \varphi}_\alpha \left( u\right) \varphi ^\alpha \left( u\right)
\right] , \n{5.5}
\e
where $f_{\alpha \beta }=D_\alpha \varphi _\beta -D_\beta \varphi _\alpha ,$
one follows the Proca equations on la-spaces
\f
D_\alpha f^{\alpha \beta }\left( u\right) +\mu ^2\varphi ^\beta \left(
u\right) =0. \n{5.6}
\e
Equations (5.6) are a first type constraints for $\beta =0.$ Acting with $%
D_\alpha $ on (5.6), for $\mu \neq 0$ we obtain second type constraints%
\f
D_\alpha \varphi ^\alpha \left( u\right) =0. \n{5.7}
\e

Putting (5.7) into (5.6) we obtain second order field equations with respect
to $\varphi _{\alpha }$ :%
\f
\Box \varphi _\alpha \left( u\right) +R_{\alpha \beta }\varphi ^\beta \left(
u\right) +\mu ^2\varphi _\alpha \left( u\right) =0. \n{5.8}
\e
The energy-momentum d-tensor and d-vector current following from the (5.5)
can be written as
$$
E_{\alpha \beta }^{(1)}\left( u\right) =-g^{\varepsilon \tau }\left(
{\overline f}_{\beta \tau } f_{\alpha \varepsilon }+
{\overline f}_{\alpha \varepsilon} f_{\beta \tau }\right) +
\mu ^2\left({\overline \varphi}_\alpha \varphi _\beta
+{\overline \varphi}_\beta \varphi _\alpha \right) -
\frac{g_{\alpha \beta }}{\sqrt{%
\left| g\right| }}{\cal L}^{(1)}\left( u\right) .
$$
and%
$$
J_\alpha ^{\left( 1\right) }\left( u\right) =
i\left( {\overline  f}_{\alpha \beta}
\left( u\right) \varphi ^\beta \left( u\right) - {\overline \varphi}^{\beta}
\left( u\right) f_{\alpha \beta }\left( u\right) \right) .
$$

For $\mu =0$ the d-tensor $f_{\alpha \beta }$ and the Lagrangian (5.5) are
invariant with respect to locally anisotropic gauge transforms of type
$$
\varphi _\alpha \left( u\right) \rightarrow \varphi _\alpha \left( u\right)
+\delta _\alpha \Lambda \left( u\right) ,
$$
where $\Lambda \left( u\right) $ is a d-differentiable scalar function, and
we obtain a locally anisotropic variant of Maxwell theory.
\vskip30pt
{\large \em 5.3. La-gravitons on la-backgrounds}
\vskip
15pt
Let a massless d-tensor field $h_{\alpha \beta }\left( u\right) $ is
interpreted as a small perturbation of the locally anisotropic background
metric d-field $g_{\alpha \beta }\left( u\right) .$ Considering, for
simplicity, a torsionless background we have locally anisotropic Fierz-Pauli
equations%
\f
\Box h_{\alpha \beta }\left( u\right) +2R_{\tau \alpha \beta \nu }\left(
u\right) ~h^{\tau \nu }\left( u\right) =0 \n{5.9}
\e
and d-gauge conditions%
\f
D_\alpha h_\beta ^\alpha \left( u\right) =0,\quad h\left( u\right) \equiv
h_\beta ^\alpha (u)=0, \n{5.10}
\e
where $R_{\tau \alpha \beta \nu }\left( u\right) $ is curvature d-tensor of
the la-background space (these formulae can be obtained by using a
perturbation formalism with respect to $h_{\alpha \beta }\left( u\right) $
developed in [27]; in our case we must take into account the distinguishing
of geometrical objects and operators on la-spaces).

We note that we can rewrite d-tensor formulas (5.1)-(5.10) into similar
d-spinor ones by using formulas (4.1)-(4.3), (4.13), (4.15) and
(4.20)-(4.29) (for simplicity, we omit these
considerations in this paper).
\vskip30pt
{\large \em 5.4. Locally anisotropic Dirac equations}
\vskip15pt
Let denote the Dirac d-spinor field on ${\cal E}$ as $\psi \left( u\right)
=\left( \psi ^{\underline{\alpha }}\left( u\right) \right) $ and consider as
the generalized Lorentz transforms the group of automorphysm of the metric $%
G_{\widehat{\alpha }\widehat{\beta }}$ (see the la-frame decomposition of
d-metric (3.1)).The d-covariant derivation of field $\psi \left( u\right) $
is written as
\f
\overrightarrow{\nabla _\alpha }\psi =\left[ \delta _\alpha +\frac 14C_{%
\widehat{\alpha }\widehat{\beta }\widehat{\gamma }}\left( u\right) ~l_\alpha
^{\widehat{\alpha }}\left( u\right) \sigma ^{\widehat{\beta }}\sigma ^{%
\widehat{\gamma }}\right] \psi , \n{5.11}
\e
where coefficients $C_{\widehat{\alpha }\widehat{\beta }\widehat{\gamma }%
}=\left( D_\gamma l_{\widehat{\alpha }}^\alpha \right) l_{\widehat{\beta }%
\alpha }l_{\widehat{\gamma }}^\gamma $ generalize for la-spaces the
corresponding Ricci coefficients on Riemannian spaces [28]. Using $\sigma $%
-objects $\sigma ^\alpha \left( u\right) $ (see (4.2)) we define the Dirac
equations on la-spaces:%
\f
(i\sigma ^\alpha \left( u\right) \overrightarrow{\nabla _\alpha }-\mu )\psi
=0, \n{5.12}
\e
which are the Euler equations for the Lagrangian%
$$
{\cal L}^{(1/2)}\left( u\right) =\sqrt{\left| g\right| } \{ [ \psi
^{+}\left( u\right) \sigma ^\alpha \left( u\right) \overrightarrow{\nabla
_\alpha }\psi \left( u\right) - $$  \f (\overrightarrow{\nabla _\alpha }\psi
^{+}\left( u\right) )\sigma ^\alpha \left( u\right) \psi \left( u\right)
 ] -\mu \psi ^{+}\left( u\right) \psi \left( u\right) \} , \n{5.13}
\e
where $\psi ^{+}\left( u\right) $ is the complex conjugation and transposition
of the column$~\psi \left( u\right) .$

From (5.13) we obtain the d-metric energy-momentum d-tensor%
$$
E_{\alpha \beta }^{(1/2)}\left( u\right) =\frac i4 [ \psi ^{+}\left(
u\right) \sigma _\alpha \left( u\right) \overrightarrow{\nabla _\beta }\psi
\left( u\right) +\psi ^{+}\left( u\right) \sigma _\beta \left( u\right)
\overrightarrow{\nabla _\alpha }\psi \left( u\right) -$$
$$ (\overrightarrow{%
\nabla _\alpha }\psi ^{+}\left( u\right) )\sigma _\beta \left( u\right) \psi
\left( u\right) -(\overrightarrow{\nabla _\beta }\psi ^{+}\left( u\right)
)\sigma _\alpha \left( u\right) \psi \left( u\right) ]
$$
and the d-vector source%
$$
J_\alpha ^{(1/2)}\left( u\right) =\psi ^{+}\left( u\right) \sigma _\alpha
\left( u\right) \psi \left( u\right) .
$$
We emphasize that la-interactions with exterior gauge fields can be
introduced by changing the la-partial derivation from (5.11) in this manner:%
\f
\delta _\alpha \rightarrow \delta _\alpha +ie^{\star }B_\alpha , \n{5.14}
\e
where $e^{\star }$ and $B_\alpha $ are respectively the constant d-vector
potential of la-gauge interactions on la-spaces (see [9] and the next
subsection).
\vskip30pt
{\large \em 5.5. D-spinor equations for Yang-Mills fields on la-spaces}
\vskip15pt
We consider a v-bundle ${\cal B}_E,~\pi _B:{\cal B\rightarrow E,}$ on
la-space ${\cal E.\,}$ Additionally to d-tensor and d-spinor indices we
shall use capital Greek letters, $\Phi ,\Upsilon ,\Xi ,\Psi ,...$ for fibre
(of this bundle) indices (see details in [12,13] for the case when the base
space of the v-bundle $\pi _B$ is a locally isotropic space-time). Let $%
\underline{\nabla }_\alpha $ be, for simplicity, a torsionless,
linear connection in ${\cal B}_E$ satisfying conditions:
$$
\underline{\nabla }_\alpha :{\em \Upsilon }^\Theta \rightarrow {\em %
\Upsilon }_\alpha ^\Theta \quad \left[ \text{or }{\em \Xi }^\Theta
\rightarrow {\em \Xi }_\alpha ^\Theta \right] ,
$$
$$
\underline{\nabla }_\alpha \left( \lambda ^\Theta +\nu ^\Theta
\right) =\underline{\nabla }_\alpha \lambda ^\Theta +%
\underline{\nabla }_\alpha \nu ^\Theta ,
$$
$$
\underline{\nabla }_\alpha ~(f\lambda ^\Theta )=\lambda ^\Theta
\underline{\nabla }_\alpha f+f\underline{\nabla }_\alpha
\lambda ^\Theta ,\quad f\in {\em \Upsilon }^\Theta ~[\mbox{or }{\em \Xi }%
^\Theta ],
$$
where by ${\em \Upsilon }^\Theta ~\left( {\em \Xi }^\Theta \right) $ we
denote the module of sections of the real (complex) v-bundle ${\cal B}_E$
provided with the abstract index $\Theta .\,\,$The curvature of connection $%
\underline{\nabla }_\alpha $ is defined as
$$
K_{\alpha \beta \Omega }^{\qquad \Theta }\lambda ^\Omega =\left(
\underline{\nabla }_\alpha \underline{\nabla }_\beta -%
\underline{\nabla }_\beta \underline{\nabla }_\alpha
\right) \lambda ^\Theta .
$$

For Yang-Mills fields as a rule one considers that ${\cal B}_E$ is enabled
with a unitary (complex) structure (complex conjugation changes mutually the
upper and lower Greek indices). It is useful to introduce instead of $%
K_{\alpha \beta \Omega }^{\qquad \Theta }$ a Hermitian matrix $F_{\alpha
\beta \Omega }^{\qquad \Theta }=i$ $K_{\alpha \beta \Omega }^{\qquad \Theta }
$ connected with components of the Yang-Mills d-vector potential $B_{\alpha
\Xi }^{\quad \Phi }$ according the formula:%
\vskip20pt
\f
\frac 12F_{\alpha \beta \Xi }^{\qquad \Phi }=\underline{\nabla }%
_{[\alpha }B_{\beta ]\Xi }^{\quad \Phi }-iB_{[\alpha |\Lambda |}^{\quad \Phi
}B_{\beta ]\Xi }^{\quad \Lambda }, \n{5.15}
\e
where the la-space indices commute with capital Greek indices. The gauge
transforms are written in the form:%

$$
B_{\alpha \Theta }^{\quad \Phi }\mapsto B_{\alpha \widehat{\Theta }}^{\quad
\widehat{\Phi }}=B_{\alpha \Theta }^{\quad \Phi }~s_\Phi ^{\quad \widehat{%
\Phi }}~q_{\widehat{\Theta }}^{\quad \Theta }+is_\Theta ^{\quad \widehat{%
\Phi }}\underline{\nabla }_\alpha ~q_{\widehat{\Theta }}^{\quad
\Theta },
$$
\vskip10pt
$$
F_{\alpha \beta \Xi }^{\qquad \Phi }\mapsto F_{\alpha \beta \widehat{\Xi }%
}^{\qquad \widehat{\Phi }}=F_{\alpha \beta \Xi }^{\qquad \Phi }s_\Phi
^{\quad \widehat{\Phi }}q_{\widehat{\Xi }}^{\quad \Xi },
$$
where matrices $s_\Phi ^{\quad \widehat{\Phi }}$ and $q_{\widehat{\Xi }%
}^{\quad \Xi }$ are mutually inverse (Hermitian conjugated in the unitary
case). The Yang-Mills equations on torsionless la-spaces [9] are written in
this form:%
\f
\underline{\nabla }^\alpha F_{\alpha \beta \Theta }^{\qquad \Psi
}=J_{\beta \ \Theta}^{\qquad \Psi}   , \n{5.16}
\e
\f
\underline{\nabla }_{[\alpha }F_{\beta \gamma ]\Theta }^{\qquad \Xi
}=0. \n{5.17}
\e
We must introduce deformations of connection of type (2.14) and (2.15),
 $\underline{\nabla }_\alpha ^{\star }~\longrightarrow
\underline{\nabla }_\alpha +P_\alpha, $ (the deformation d-tensor $%
P_\alpha $ is induced by the torsion in v-bundle ${\cal B}_E)$ into the
definition of the curvature of la-gauge fields (5.15) and motion equations
(5.16) and (5.17) if interactions are modeled on a generic la-space.

Now we can write out the field equations of the Einstein-Cartan theory in
the d-spinor form. So, for the Einstein equations (2.42) we have
\vskip15pt
$$
\overleftarrow{G}_{\underline{\gamma }_1\underline{\gamma }_2\underline{%
\alpha }_1\underline{\alpha }_2}+\lambda \varepsilon _{\underline{\gamma }_1%
\underline{\alpha }_1}\varepsilon _{\underline{\gamma }_2\underline{\alpha }%
_2}=\kappa E_{\underline{\gamma }_1\underline{\gamma }_2\underline{\alpha }_1%
\underline{\alpha }_2},
$$
with $\overleftarrow{G}_{\underline{\gamma }_1\underline{\gamma }_2%
\underline{\alpha }_1\underline{\alpha }_2}$ from (4.22), or, by using the
d-tensor (4.23),%
\vskip15pt
$$
\Phi _{\underline{\gamma }_1\underline{\gamma }_2\underline{\alpha }_1%
\underline{\alpha }_2}+(\frac{\overleftarrow{ R}}4-
\frac \lambda 2)\varepsilon _{\underline{%
\gamma }_1\underline{\alpha }_1}\varepsilon _{\underline{\gamma }_2%
\underline{\alpha }_2}=-\frac \kappa 2E_{\underline{\gamma }_1\underline{%
\gamma }_2\underline{\alpha }_1\underline{\alpha }_2},
$$
which are the d-spinor equivalent of the equations (2.44). These equations
 must be completed by the algebraic equations
(2.45) for the d-torsion and d-spin density with d-tensor indices changed
into corresponding d-spinor ones.
\vskip50pt
{\large \bf 6. Outlook  and discussion}
\vskip15pt
In summary, we have developed the spinor differential geometry of vector
bundles provided with nonlinear and distinguished connections and metric
structures and shown in detail the way of formulation the theory of
fundamental field (gravitational, gauge and spinor) interactions on generic
 locally anisotropic spaces.

Let us draw attention to some evidences for the necessity to take into account
physical effects of possible local anisotropy. The first is that from modern
higher dimensional field theories being  low-energy limits of the string
theory. In this case a {\bf nonlinear connection} (2.3) can be considered as
a specific field describing the dynamical splitting of the high dimensional
space into, for instance, four dimensional space-time and the rest of dimensions
one. We can impose different type of field equations for components of a
N-connection and its curvature in order to modelate different scales of
local anisotropy at macroscopic or microscopic levels (for instance, in
 [10] we define such equations as some constraints necessary for consistent
 propagation of strings in a background la-space; without the loss of
generality we have not imposed in this work any restrictions on the type of
 N-connection
structure). Kaluza-Klein models are obtained for trivial N-connections
and by neglecting the dependence on "relic" higher dimension $y$-variables
of physical objects on the four dimensional space-time. As a second evidence
 can be considered the well known result that a self-consistent theoretical
description of radiational processes in classical field theories is possible
by adding high derivation terms. The third evidence for the necessity of
 investigation of quantum models on tangent and vector bundles (with both
coordinate and momenta type variables) is the general problem of the status of
singularities and of renormalization in quantum field theory. It is also
clear that a careful analysis of physical processes when the back reaction
of classical and quantum systems interacting or being measured is not
negligible small, or investigation, for instance, of gravitational radiational
dissipation in all variants of classical  and quantum gravity and quantum
field theories with high derivatives require extensions of  geometrical
 backgrounds of theories.

Of course, the interpretation and physical consequences of the N-connection structure
are different in various models with local anisotropy (models of continuous
media with dislocations and disclinations, locally anisotropic stochastic
processes with turbulent space-time, curved momentum, or phase spaces, or
cosmological models with global and local anisotropy). The main advantages
of the approach to the geometry of la-spaces developed by R. Miron and
M. Anastasiei [2] are the generality and admissibility of a differential
and integral calculus adapted to the N-connection structure. We emphasize
that this way, in locally adapted bases, one holds a lot of geometric
 similarities with locally isotropic Einstein-Weyl-Cartan spaces (see for
details the mentioned monograph and the presented there references
and, in brief, the second section of this paper). More than that, if
the first physical  models based on Finsler geometry
 were rather sophisticate, with ambiguities and characterized by tedious
 calculations, the general modeling of la-spaces
as vector bundles provided with compatible N-connection, d-connection and
metric structures,  make it possible to formulate  theories of classical
and quantum field locally anisotropic interactions in a manner very similar
to that for locally isotropic curved spaces.

There were two main impediments to the developing of physical theories on
 la-spaces: the problem of formulation of conservation laws on such spaces
and the definition of la-spinors. It is well known, for instance, that the
conservation of the energy-momentum values is a consequence of the invariance
of the fundamental space-time under the action of it global group of
automorphysms (for the Minkowski space one considers the Poincare group).
In general relativity we have only local Lorentz rotations and in consequence
there are nontrivial questions with the definition of energy-momentum type
 values for gravitational field (as attempts to solve this problem one
introduces pseudo-tensor objects or two metric structures [29], formulates
the Einstein theory on arbitrary space-times [27], or uses nearly
autoparallel maps of curved spaces [30]). On generic la-spaces we are not
able to introduce groups of automorphysms even locally and, at first glance,
it is impossible to formulate any type of conservation laws and to define
spinor objects, i.e., in consequence, we can not propose any general
principles of formulation physical models with la-interactions in a consequent
manner. Nevertheless, it was happen that variants of solution of such problems
 are possible by modeling la-spaces on v-bundles (on conservation laws in
Lagrange spaces see [18] and the paper by Vacaru S and Ostaf S in Ref. [3],
as well [19], for generalizations to arbitrary la-spaces; the problem of
definition of locally anisotropic spinor and twistors is studied in [7,8,23,
21]).

The distinguishing by a N-connection structure of the multidimensional space
into horizontal and vertical subbundles points out to the necessity to
start up the spinor constructions for la-spaces with a study of distinguished
Clifford algebras for vector spaces split into h- and v-subspaces. The
d-spinor objects exhibit a eight-fold periodicity on dimensions of the
mentioned subspaces. As it was shown in [7,8], see also the section 3 of this
work, a  corresponding d-spinor technique can be developed,
which is very similar to that proposed by Penrose and Rindler [11-13]
for locally isotropic curved spaces if the locally adapted to the
N-connection structures d-spinor and d-vector frames are used. It is clear
the d-spinor calculus is more tedious than the 2-spinor
one for Einstein spaces because of multidimensional and multiconnection
character of generic la-spaces.

The d-spinor differential geometry formulated in the section 4 can be
considered as a branch of the geometry of Clifford fibrations for v-bundles
provided with
N-connection, d-connection and metric structures (the necessary geometric
background can be found in [7,20]). We have emphasized only the features
containing d-spinor torsions and curvatures which are necessary for
 a d-spinor formulation of la-gravity. To develop a conformally invariant
d-spinor calculus is possible only for a particular class of la-spaces when
the Weyl d-tensor (2.40) is defined by the N-connection and d-metric
structures. In general, we have to extend the class of conformal
transforms to that of nearly autoparallel maps of la-spaces [19,21,23,31].
This is a matter of our further investigations.

Having fixed compatible N-connection, d-connection and metric structures on
a la-space $\cal E$ we can develop physical models on this space by using
a covariant variational d-tensor calculus as on Riemann-Cartan spaces (really
there are specific complexities because the d-torsion in general is not
antisymmetric, the Ricci d-tensor is not symmetric and the la-frames are
nonholonomic). The systems of basic field equations for fundamental matter
(scalar, Proca and Dirac) fields and gauge and gravitational fields have been
introduced in a geometric manner by using d-covariant operators and la-frame
decompositions of d-metric. These equations and  expressions
for energy-momentum d-tensors and d-vector currents can be established by
using the standard variational procedure, but correspondingly adapted to the
N-connection structure if we work by using la-bases.

Finally, we should note that d-spinor technique developed in this paper
can be applied in modern string and superstring theories [10], quantum
gravity and cosmology and gauge like approaches to fundamental interactions
when the multidimensional and/or locally anisotropic, radiational and
stochastic processes play an important role and have significant
physical consequences.

\acknowledgments

The author warmly thank Profs R. Miron and M. Anastasiei for helpful
discussions. A part of this work was completed when he was visiting
Moscow and enjoying the support and hospitality of Dr E. Seleznev
and Ms T. Kushnareva.

\end{document}